\documentclass{xjkas}

\def\year{2015} 
\def\month{October} 
\def\volume{00} 
\def\beginpage{1} 
\def\endpage{14} 
\def\received{July 27, 2015} 
\def\accepted{October 25, 2015} 
\date{Received \received; accepted \accepted}

\usepackage{color}
\usepackage[breaklinks, colorlinks, citecolor=blue, linkcolor=blue, menucolor=blue, urlcolor=blue]{hyperref}

\title{
PAGaN I: Multi-Frequency Polarimetry of AGN Jets with KVN\thanks{Part of a special issue on the Korean VLBI Network (KVN)}
}

\author[1,2]{Jae-Young~Kim}
\author[1]{Sascha~Trippe}
\author[3]{Bong~Won~Sohn}
\author[1]{Junghwan~Oh}
\author[1]{Jong-Ho~Park}
\author[3]{Sang-Sung~Lee}
\author[1]{Taeseok~Lee}
\author[1]{Daewon~Kim}

\affil[1]{Department of Physics and Astronomy, Seoul National University, Seoul 08826, Korea; \email{trippe@astro.snu.ac.kr}}
\affil[2]{Max-Planck Institut f\"{u}r Radioastronomie, Auf dem H\"{u}gel 69, D-53121 Bonn, Germany; \email{jykim@mpifr-bonn.mpg.de}}
\affil[3]{Korea Astronomy and Space Science Institute, Daejeon 34055, Korea}

\def\corrauthor{
S. Trippe
}

\def\runningauthor{
Kim et al.
}

\def\runningtitle{
PAGaN I: Multi-Frequency Polarimetry of AGN Jets with KVN
}

\def\keywords{
galaxies: active, jets -- radio continuum: galaxies -- techniques: interferometric, polarimetric
}

\def\abstracttext{
Active Galactic Nuclei (AGN) with bright radio jets offer the opportunity to study the structure of and physical conditions in relativistic outflows. 
For such studies, multi-frequency polarimetric very long baseline interferometric (VLBI) observations are important as they directly probe particle densities, magnetic field geometries, and several other parameters. 
We present results from first-epoch data obtained by the Korean VLBI Network (KVN) within the frame of the \emph{P}lasma Physics of \emph{A}ctive \emph{Ga}lactic \emph{N}uclei (PAGaN) project. 
We observed seven radio-bright nearby AGN at frequencies of 22, 43, 86, and 129 GHz in dual polarization mode.  
Our observations constrain apparent brightness temperatures of jet components and radio cores in our sample to $>10^{8.01}$\,K and $>10^{9.86}$\,K, respectively.
Degrees of linear polarization $m_{L}$ are relatively low overall: less than 10\%. 
This indicates suppression of polarization by strong turbulence in the jets. 
We found an exceptionally high degree of polarization in a jet component of BL Lac at 43 GHz, with $m_{L} \sim$ 40\%. 
Assuming a transverse shock front propagating downstream along the jet, the shock front being almost parallel to the line of sight can explain the high degree of polarization.
}

\begin{document}
\jkashead 


\section{Introduction\label{sec:intro}}

Active Galactic Nuclei (AGN) are the most energetic persistent sources of electromagnetic radiation in the universe, powered by accretion of matters onto central supermassive black holes (SMBHs) with masses $M_{\bullet} \sim 10^{6} - 10^{9} M_{\odot}$. AGN with relativistic radio jets are of particular interest because of their enormous internal energies and complicated outflow structures on a wide range of spatial scales, from sub-parsec to kiloparsec scales.
There is general consensus that jets originate from magnetohydrodynamical interaction between accreted matter and magnetic fields in the immediate environment (few Schwarzschild radii) of rotating black holes. However, details such as launching, acceleration, and collimation processes remain poorly understood (see, e.g., \citealt{boettcher12} for a review).

A better understanding of the physics and evolution of jet outflows requires observations of 
(i) the internal structure of the plasma, such as particle density distributions, and 
(ii) the magnetic fields pervading the outflows. Technically, this can be achieved via multi-frequency polarimetric radio observations of the synchrotron radiation from relativistic electrons in the jet plasma -- in particular, via high-resolution polarimetric imaging observations with very long baseline interferometry (VLBI).
VLBI observations provide a three-dimensional view on matter distributions and on the magnetic field configuration inside jets \citep[e.g.,][]{walker00,asada02,kadler04,lister05,gomez08,osullivan09a,osullivan09b,molina14, kim14,park15}.
Time-series analysis of spectra and/or polarization, combined with kinematic information, has revealed the evolution of internal shocks in jets  \citep{jorstad05,jorstad07,trippe10,trippe12b,fromm13,park14}.

Even though a lot of progress has been made at centimeter wavelengths, multi-frequency polarimetric studies at millimeter wavelengths are still rare. 
Such observations are important because they are less affected by synchrotron self-absorption than observations at cm wavelengths and thus probe emission from the inner regions of jets. 
Recent 86-GHz VLBI surveys of compact radio jets (e.g. \citealt{lee08,lee14b}) revealed that the brightness temperatures of cores are systematically lower at mm wavelengths than at cm wavelengths, which is indicative of kinetic-energy dominated outflows around jet bases.
Rotation measure (RM) analyses at mm-wavelengths performed by \cite{marrone06}, \cite{trippe12b}, \cite{plambeck14}, and \cite{kuo14} found very high RM values on the order of $10^{4-5}$\,rad\,$\rm m^{-2}$. 
Such high values translate into strong magnetic fields and high electron densities characteristic of accretion processes near the central black holes.

As illustrated above, mm-VLBI polarimetry at multiple frequencies is required to provide a fresh view on the plasma physics of the jets in active galaxies. Due to its wide frequency coverage (22--129~GHz) and the capability to observe at up to four frequencies simultaneously, the Korean VLBI Network (KVN; \citealt{lee14a}) is an excellent tool for probing jet plasmas at mm wavelengths.
In addition, the KVN and VERA Array (KaVA; \citealt{niinuma14}) provides the good $uv$ coverage needed for detailed investigations of jet structure and kinematics.
With this in mind, we initiated the \emph{P}lasma Physics of \emph{A}ctive \emph{Ga}lactic \emph{N}uclei (PAGaN) project, a systematic radio interferometric study of AGN radio jets.

In this paper, we present results from our first-epoch polarimetric observations obtained by KVN. In a companion paper (PAGaN II, \citealt{oh2015}) we discuss first results from KaVA. Throughout our paper, we adopt a cosmology with $H_{0}=71$ km\,s$^{-1}$\,Mpc$^{-1}$, $\Omega_{\Lambda}=0.73$, and $\Omega_{\rm m}=0.27$.

\section{Observations and Analysis\label{sec:method}}

\subsection{Observations and Data Reduction\label{sec:reduction}}

\begin{table}[t!]
\caption{Our target sources.\label{tab:list}}
\centering
\begin{tabular}{lrrr}
\toprule
Name$^{\rm a}$ & Type$^{\rm b}$ & Redshift & Scale$^{\rm c}$ \\
&&& [pc/mas] \\ 
\midrule
3C 111 (0415+379) & RG & 0.0491 & 0.95 \\
3C 120 (0430+052) & RG & 0.033  & 0.65 \\
3C 84  (0316+413) & RG & 0.0176 & 0.30 \\
4C +01.28  (1055+018) & Q & 0.888 & 7.79 \\
4C +69.21  (1642+690) & Q & 0.751 & 7.35 \\
BL Lac  (2200+420) & BL & 0.0686 & 1.29 \\
DA 55 (0133+476) & Q & 0.859 & 7.70 \\
\bottomrule
\end{tabular}
\tabnote{
Basic source information from the MOJAVE database.
(a) Conventional source names, followed by corresponding B1950 identifiers.
(b) Classification; RG: Radio Galaxy; Q: quasar; BL: BL Lacertae object.
(c) Image scale for a given redshift, in pc/mas.
}
\end{table}

For our observations, we focused on AGN that (1) were sufficiently bright, with total fluxes $\gtrsim1$\,Jy at 22 GHz; (2) showed extended ($\gtrsim10$\,mas) structure; and, as far as known, (3) showed significant linear polarization.
We referred to the Very Long Baseline Array (VLBA) monitoring database maintained by 
the MOJAVE program\footnote{\url{http://www.physics.purdue.edu/astro/MOJAVE/}} and
VLBA-BU-BLAZAR program\footnote{\url{http://www.bu.edu/blazars/VLBAproject.html}}
in order to check total fluxes, structures, and linear polarizations at 15 and 43 GHz, respectively.
Based on the source information from the databases we selected seven sources:
3C 111, 3C 120, 4C +01.28, 4C +69.21, and BL Lac, all known for spatially extended and linearly polarized structure; 
3C 84, known for its high radio brightness ($\gtrsim30$\,Jy at 22 GHz) and its usefulness as polarization calibrator; 
and DA 55, known for its relatively high radio brightness ($\gtrsim$5\,Jy at 22 GHz) and  high polarized flux ($\sim$200\,mJy at 43 GHz).
A brief summary of our targets is given in Table \ref{tab:list}.

The selected sources were observed in 15 sessions between 21 October 2013 and 9 March 2014.
Each source was observed over a full track to ensure high sensitivity and good $uv$ coverage.
In addition, we obtained multiple `snapshot' observations resulting from scheduling constraints.
Observing frequencies were 21.7, 42.9, 86.6, and 129.9 GHz, respectively. At each frequency, the bandwidth was 64 MHz (16 MHz $\times$ 4 IFs) and both left and right circular polarizations (LCP/RCP) were observed.
Thanks to the unique capabilities of KVN, we obtained simultaneous dual polarization observations at two pairs of frequencies (either 22/43 GHz or 86/129 GHz).
The observed data were correlated at the Korea-Japan Correlation Center (KJCC; \citealt{lee2015}).

\begin{figure}[t!]
  \centering
  \includegraphics[trim=2mm 2mm 7mm 10mm, clip, width=80mm]{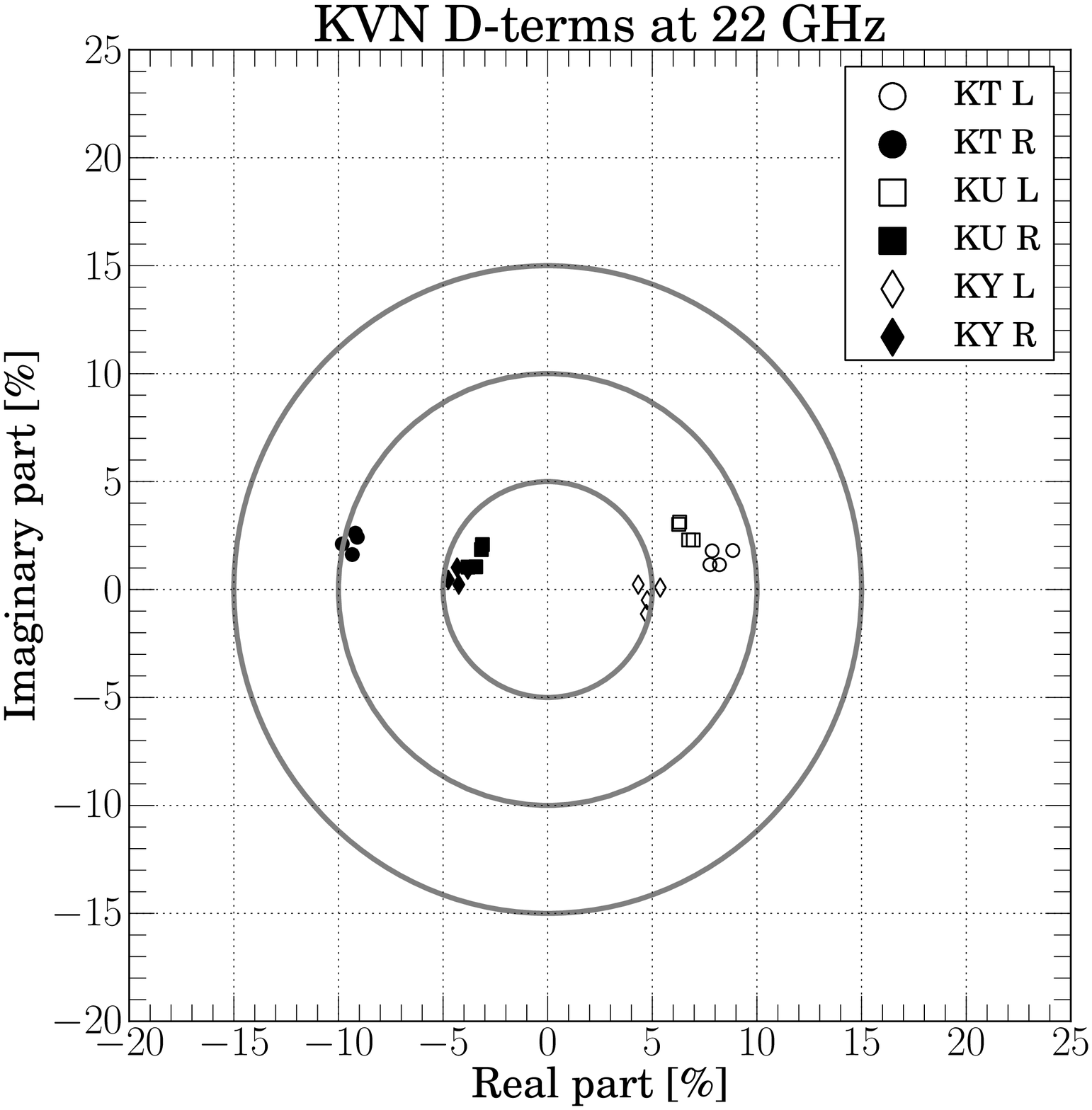}
  \includegraphics[trim=2mm 2mm 7mm 10mm, clip, width=80mm]{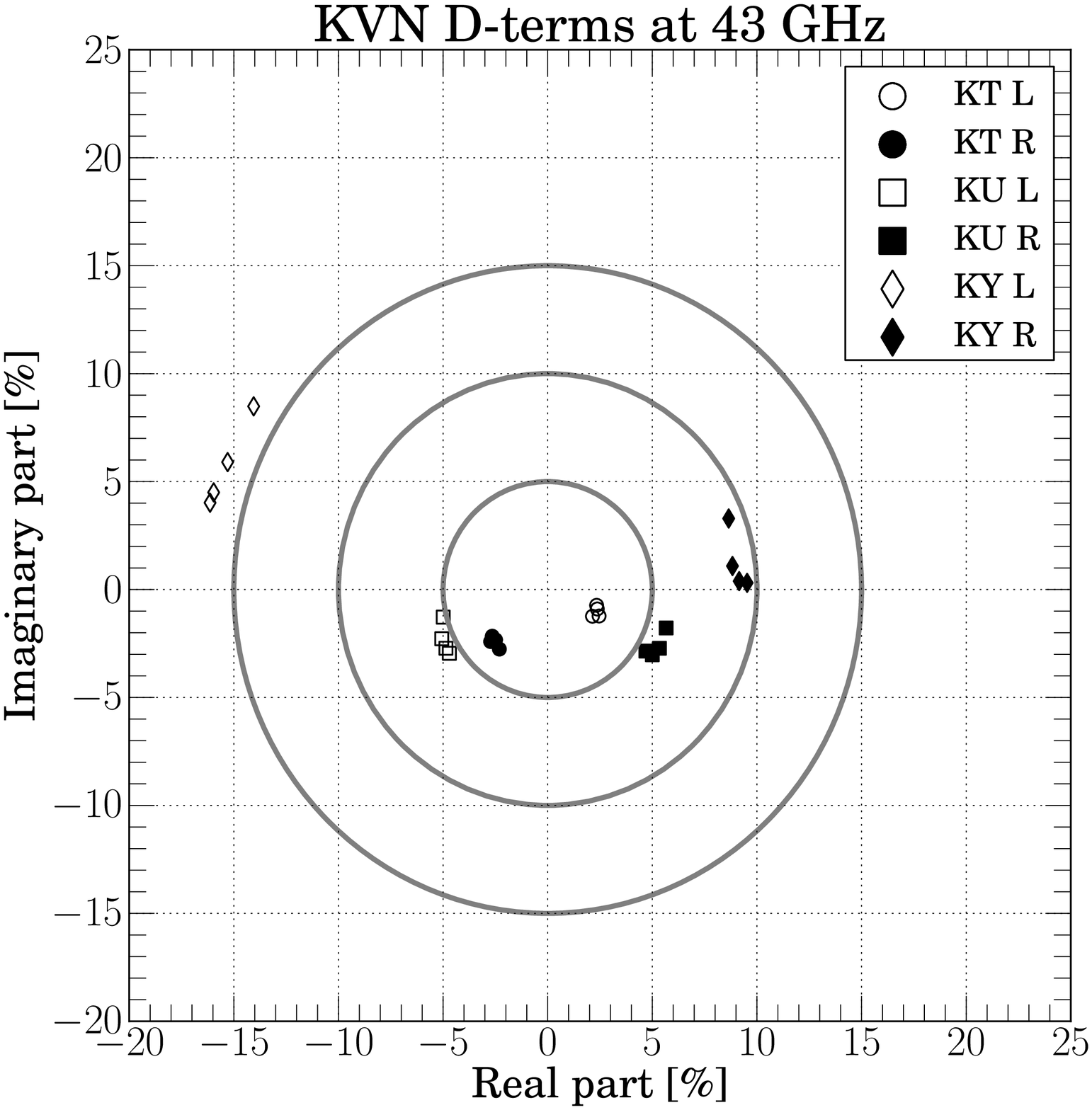}
  \caption{D-terms of the KVN antennas.
	 \textit{Top:} The D-terms of KT, KU, and KY stations at 22 GHz. L and R denote LCP and RCP receivers, respectively. Four D-terms are obtained for four IFs at each station and each polarization. Grey circles indicate 5\%, 10\%, and 15\% contours.
	 \textit{Bottom:} Same as top panel, at 43 GHz.
	 The D-terms of the 4 IFs at each frequency and each polarization have similar absolute values and phases, indicating good quality and stability of the instrumental polarization calibration.
  }
  \label{fig:Dterms}
\end{figure}

We processed the correlated visibility data with the software package \textsf{AIPS}\footnote{Provided and maintained by the National Radio Astronomical Observatory of the U.S.A.} to perform standard calibration steps such as fringe fitting, amplitude calibration, and bandpass correction.
As KVN has only three baselines, the polarimetric calibration needed to be performed carefully.
We mapped the targets with \textsf{Difmap} by means of CLEANing and phase self-calibration with natural weighting (i.e., \textit{uvweight} 0, $-2$).
The CLEANed total intensity maps were used for amplitude self-calibration of the L and R visibilities 
(using the task \textsf{AIPS CALIB} with \textit{A\&P} and \textit{L1R} options),
and for determining the instrumental polarization leakages (D-terms, with the task \textsf{LPCAL}) for individual IF.
For most observing sessions, the unpolarized radio galaxy 3C 84 served as D-term calibrator.
We examined the D-terms using the methods described in \cite{roberts94} and \cite{aaron97} (i.e. calculating cross- to parallel-hand visibility ratio before/after applying the D-term corrections)
and found that those D-terms derived successfully are reliable up to frequencies of 43 GHz.
We show the D-terms at 22 and 43 GHz obtained from one of our observing sessions in Figure~\ref{fig:Dterms} for illustration.

We note that our unsuccessful polarimetric observations for several sources and several frequencies could have suffered from
(i) imperfect calibration of instrumental L and R gain offsets, and (ii) poor quality of estimated D-terms.
The gain calibration is rather unlikely to have affected our results because we found that the LL and RR dirty maps of calibrated sources have the same peak values within their uncertainties $\sim$1\% and the peak values of the dirty Stokes V maps are typically less than a few mJy.
Therefore we expect that the quality of the D-terms is the more important factor in general.
For instance, we had limited $uv$ coverage for the D-term calibrator (i.e. 3C 84) in several observing runs due to scheduling constraints, leading to \textsf{LPCAL} not producing suitable solutions.
Poorly determined D-terms leave residual instrumental polarization in the cross-hand visibilities and those residuals could have contributed to amplitude errors or phase offsets in Stokes Q and U maps.

Calibration of the absolute electric vector position angles (EVPAs) is important because D-terms correct for only the polarization leakages in the antennas and the phase self-calibration does not preserve source-intrinsic relative phase of cross-hand visibilities.
The absolute EVPAs of target sources can be recovered when a calibrator with well-known stable EVPA is simultaneously observed with the array in VLBI mode.
Unfortunately, we failed to calibrate the EVPAs because the quasar 3C~286, which was intended to serve as calibrator source, was not detected in our data.
Therefore only relative EVPA information is available for the data in this paper.
The EVPA calibration can be accomplished in another way if any strongly polarized compact source is observed in both single-dish and VLBI mode (quasi-) simultaneously under the assumption that EVPA of the calibrator is the same on both scales.
This method may be useful for future KVN polarimetry as long as no bright calibrators with stable EVPAs are available at mm-wavelengths.

\subsection{Data Analysis\label{sec:analysis}}

\begin{table*}[t!]
\centering
\caption{Flux densities, spectral indices, and degrees of polarization of our targets, integrated over all source components.}
\begin{tabular}{lcccccc}
\toprule
Source & Epoch & $\nu_{\rm obs}$ & $S_{\rm tot}$ & $\alpha$ & $ S_{L} $ & $ m_{L} $ \\
(1)&  (2)  &  (3)  &      (4)      &    (5)   &       (6)       &       (7)       \\
\midrule
3C 111 (0415+379) &  2014/01/03 & 22 & 3.56 $\pm$ 0.17 & 0.14 $\pm$ 0.13  &   42.0 $\pm$ 27.0 &  1.2 $\pm$ 0.8  \\
                  &  2014/01/03 & 43 & 3.92 $\pm$ 0.28 &                  &                   &        \\
                  &  2014/03/06 & 86 & 2.16 $\pm$ 0.27 &$-$0.38 $\pm$ 0.41  &                   &        \\
                  &  2014/03/06 &129 & 1.85 $\pm$ 0.21 &                  &                   &        \\
\midrule  
3C 120 (0430+052) & 2013/10/21 & 22 & 2.05 $\pm$ 0.07 & 0.25 $\pm$ 0.11  &   29.0 $\pm$ 9.0 &  1.4 $\pm$ 0.4   \\
                  & 2013/10/21 & 43 & 2.44 $\pm$ 0.16 &                  &                   &        \\
                  & 2014/02/16 & 86 & 1.57 $\pm$ 0.17 &                  &                   &        \\
                  & 2014/02/16 &129 &                 &                  &                   &        \\
\midrule  
3C 84 (0316+413)  &  2013/10/21 & 22 & 29.3 $\pm$ 1.70 &$-$0.46 $\pm$ 0.11  &                   &        \\
                  &  2013/10/21 & 43 & 21.4 $\pm$ 1.13 &                  &                   &        \\
                  &  2014/02/16 & 86 & 12.6 $\pm$ 0.77 &  &                   &        \\
                  &  2014/02/16 &129 &  &                  &                   &        \\
\midrule  
4C+01.28(1055+018)& 2013/10/25 & 22 & 2.52 $\pm$ 0.20 &$-$0.01 $\pm$ 0.19  &                   &        \\
                  & 2013/10/25 & 43 & 2.50 $\pm$ 0.26 &                  &                   &        \\
                  & 2014/03/09 & 86 & 2.62 $\pm$ 0.28 &$-$0.58 $\pm$ 0.54  &                   &        \\
                  & 2014/03/09 &129 & 2.24 $\pm$ 0.48 &                  &                   &        \\
\midrule  
BL Lac (2200+420) &  2014/01/02 & 22 & 5.65 $\pm$ 0.38 & 0.07 $\pm$ 0.15  &  72.3 $\pm$ 95.9  &  1.3 $\pm$ 1.7   \\
                  &  2014/01/02 & 43 & 5.92 $\pm$ 0.44 &                  &   469 $\pm$ 144   &  7.9 $\pm$ 2.5   \\
                  &  2014/02/16 & 86 & 2.27 $\pm$ 0.29 &                  &                   &        \\
                  &  2014/02/16 &129 &  &                  &                   &        \\
\midrule  
DA 55  (0133+476) &  2013/10/22 & 22 & 4.65 $\pm$ 0.16 &$-$0.13 $\pm$ 0.07  &  161 $\pm$ 39.1   &  3.5 $\pm$ 0.8   \\
                  &  2013/10/22 & 43 & 4.26 $\pm$ 0.13 &                  &  170 $\pm$ 56.2   &  4.0 $\pm$ 1.3   \\
                  &  2014/03/05 & 86 & 1.61 $\pm$ 0.35 &$-$1.42 $\pm$ 0.80  &                   &        \\
                  &  2014/03/05 &129 & 0.90 $\pm$ 0.22 &                  &                   &        \\
\bottomrule
\end{tabular}
\tabnote{
Columns: 
(1) Source name (with B1950 label).
(2) Observing epoch (in yyyy/mm/dd).
(3) Observing frequency (in GHz).
(4) Flux densities, integrated over the map (in Jy).
(5) Spectral index, average of the values derived from pairs of adjacent frequencies (i.e., $\alpha_{\rm 22,43 GHz}$ and $\alpha_{\rm 86, 129GHz}$).
(6) Polarized flux density, integrated over the source map (in mJy).
(7) Degree of polarization following from column (4) and (6) (in \%).
Blanks entries: calibration failed due to high system temperature, including all observations of 4C+69.21 (not listed here).
}
\label{tab:map_integrated}
\end{table*}

\begin{table*}[!t]
\centering
\caption{Properties of the total and polarized intensity maps for each source and each frequency.}
\begin{tabular}{l c c c c c c c c}
\toprule
$\nu_{\rm obs}$ & bmin & bmaj & bpa & Pixel size & RMS & Peak & RMS (pol) & Peak (pol) \\ 
{[GHz]} & [mas] & [mas] & [$^{\circ}$ N$\rightarrow$E]& [mas/pixel] & [mJy/beam] & [Jy/beam] & [mJy/beam] & [mJy/beam] \\
\midrule
\textbf{3C 111}   &      &       &       &      &      &     & \\
22  & 3.13 & 5.17  & $-$83.7 & 0.20 & 4.29 & 3.27 & 2.5 & 23.0 \\
43  & 1.61 & 2.62  & $-$82.6 & 0.10 & 10.3 & 3.60 &     &      \\
86	 & 0.77 & 1.31  & $-$83.2 & 0.10 & 17.4 & 1.62 &     &  	   \\
129 & 0.51 & 0.91  & $-$69.1 & 0.05 & 12.0 & 1.19 &	  &      \\
\midrule
\textbf{3C 120}   &      &       &       &      &      &     & \\
22 & 3.37 & 5.26  & $-$83.9 & 0.20 & 1.30 & 1.85 &  1.6 & 15.1 \\
43 & 1.74 & 2.65  & $-$80.5 & 0.10 & 5.15 & 2.17 &      &\\
86	& 0.82 & 1.38  & $-$77.2 & 0.10 & 9.06 & 1.19 &      &\\
\midrule
\textbf{3C 84} &  &      &       &       &      &      &     &   \\
22 & 2.75 & 5.71  & 84.9 & 0.20 & 48.2 & 22.4  & &\\
43 & 1.42 & 2.95  & 85.6 & 0.10 & 29.2 & 11.8  & &\\
86	& 0.74 & 1.30  & 84.1 & 0.10 & 22.7 & 5.86  & &\\
\midrule
\textbf{4C +01.28} &  &      &       &       &      &      &    &  \\
22 & 3.56 & 5.47  &  82.3 & 0.20 & 7.74 & 2.45 & &\\
43 & 1.81 & 2.75  &  81.7 & 0.10 & 13.7 & 2.36 & &\\
86	& 0.92 & 1.32  & $-$88.4 & 0.10 & 14.5 & 2.62 & &\\
129& 0.56 & 0.94  & $-$65.2 & 0.10 & 48.7 & 1.90 & &\\
\midrule
\textbf{BL Lac} &  &      &       &       &      &      &    &  \\
22 & 3.07 & 5.48  & $-$71.0 & 0.20 & 12.4 & 5.18 & 15.0 & 162.5 \\
43 & 1.56 & 2.70  & $-$75.2 & 0.10 & 15.7 & 5.11 & 10.9 & 154.3 \\
86	& 0.77 & 1.74  &  66.1 & 0.10 & 18.0 & 1.69 & &\\
\midrule
\textbf{DA 55} &  &      &       &       &      &      &    &  \\
22 & 3.19 & 5.24  & $-$87.5 & 0.20 & 2.64 & 4.66 & 9.4  & 135.9 \\
43 & 1.62 & 2.65  & $-$87.4 & 0.10 & 2.11 & 4.17 & 8.4 & 166.1 \\
86	& 0.75 & 1.39  & $-$76.0 & 0.10 & 37.0 & 1.69 & &\\
129& 0.57 & 0.94  & $-$65.2 & 0.05 & 24.4 & 1.90 & &\\
\bottomrule
\end{tabular}
\tabnote{
Parameters: bmin, bmaj, and bpa are the minor axis, major axis, and position angles of the clean beams, respectively.
Peak and RMS are the peak flux and rms levels of the total and polarized (`pol') intensity maps, respectively.
Blank entries: no data obtained due to failed instrumental polarization calibration.
}
\label{tab:img_params}
\end{table*}

We generated Stokes I, Q, and U maps using \textsf{Difmap} with natural weighting.
To extract source information, we fit circular Gaussian components to the calibrated $uv$ data using the \textsf{modelfit} task.
Noise (r.m.s.) levels were estimated from the residual maps.
For a given Gaussian component, we obtained 
(1) the total flux density $S_{\rm tot}$ in Jy, 
(2) the peak intensity $I_{\rm peak}$ in Jy/beam, 
(3) the map rms level $\sigma_{\rm rms}$ in mJy/beam, 
(4) the component size $d$ in mas, 
(5) radial distance of the component from the phase center of a map, $r$, in mas,
and
(6) the position angle of the component measured from north to east, $\theta$, in mas.
Following \cite{fomalont99}, the uncertainties of these parameters are 
\begin{eqnarray}
\sigma_{\rm peak} &=&   \sigma_{\rm rms}\left(1+ \frac{S_{\rm peak}}{\sigma_{\rm rms}}\right)^{1/2}
\label{eq:errors1} \\
\sigma_{\rm tot} &=&   \sigma_{\rm peak}\left(1+ \frac{S^{2}_{\rm tot}}{S^{2}_{\rm peak}} \right)^{1/2}
\label{eq:errors2} \\
\sigma_{d} &=& d\,\frac{\sigma_{\rm peak}}{S_{\rm peak}} 
\label{eq:errors3} \\
\sigma_{r} &=& \frac{1}{2} \sigma_{d}
\label{eq:errors4} \\
\sigma_{\theta} &=&  \arctan\left(\frac{\sigma_{r}}{r} \right) ~.
\label{eq:errors5} 
\end{eqnarray}

For naturally weighted $uv$ data, the minimum resolvable angular size $d_{\rm min}$ of a Gaussian component depends on the signal-to-noise ratio  $\textrm{SNR}=S_{\rm tot}/\sigma_{\rm tot}$ like
\begin{equation}
d_{\rm min} = \psi_{\rm beam} \sqrt{ \frac{4 \ln{2}}{\pi} \ln\left(\frac{\rm SNR}{\rm SNR -1}\right)    } 
\label{eq:min_size}
\end{equation}
\citep{lobanov05} where $\psi_{\rm beam}$ is the CLEAN beam size.
The above relation can be used to find the apparent brightness temperature $T_{\rm b,app}$.
At radio wavelengths, $T_{\rm b,app}$ can be expressed as
\begin{equation}
T_{\rm b,app} = 1.221\times 10^{12}\,\frac{S_{\rm tot}}{d^{2}\,\nu^{2}_{\rm obs}}(1+z) ~ \rm K \\
\label{eq:tb}
\end{equation}
where $S_{\rm tot}$ is the flux density in Jy, $\nu_{\rm obs}$ is the observing frequency in GHz, and $z$ is the redshift. The component size $d$, in mas, is either the measured size or $d_{\rm min}$ in Equation~(\ref{eq:min_size}) if the measurement returns a component size smaller than $d_{\rm min}$ (i.e., if the measured component size is consistent with zero within errors). We note that this happened at multiple occasions throughout our analysis of Stokes $I$ maps; this appears to be characteristic for KVN data (which are marked by sparse $uv$ coverage). In addition, the situation was more complicated when dealing with Stokes $Q$ and $U$ maps since no self-calibration based on Stokes $Q$ and $U$ CLEAN components could be applied to the cross-hand visibilities.

After processing the source maps for each frequency and each Stokes parameter, we extracted polarimetric and spectral properties of individual emission features by cross-referencing the maps.
Extracting the polarized flux $p$ of the $i$th jet component, $p_{i}=\sqrt{Q_{i}^{2}+U_{i}^{2}}$, and the corresponding degree of linear polarization $m_{i,L}=p_{i}/S_{i,\rm tot}$ requires the components in the I, Q, and U maps to be matched spatially. We cross-identified components via visual inspection as follows. 
\begin{itemize}

\item Adopt the center position of a jet component in the Stokes I map as a reference. 

\item Find corresponding model-fit emission features around the fixed reference position in Stokes Q and U maps.

\item Identify Q and U components located less than half the beam size away from the reference position as counterparts of the Stokes I component.

\item For the three I, Q, and U components, determine the mean and the standard deviation of the positions as true position and associated error.

\end{itemize}

We computed the spectral index of the $i$th jet component like $\alpha=\log(S_{i, \nu_{2}}/S_{i, \nu_{1}})/\log(\nu_{2}/\nu_{1})$ (i.e., $\alpha<0$ for steep spectra)
from the fluxes of components in two maps of different frequencies. We also produced two-dimensional total intensity, polarized intensity, and relative polarization angle maps using the \textsf{AIPS COMB} task.

\section{Results\label{sec:results}}

In the following, we present observational results for six AGN at multiple frequencies.
Table \ref{tab:map_integrated} summarizes observing dates, integrated fluxes, spectral indices, and vector-summed polarization properties of the sources.
Typical rms noise levels in the source maps were a few to 10 mJy/beam, with beam sizes of about 4, 2, 1, and 0.7 mas at 22, 43, 86, and 129 GHz, respectively.
Flux measurements succeeded for most of the maps except for those of 3C~120, 3C~84, and BL Lac at 129 GHz which suffered from 
high system temperatures (400~K$<T_{\rm sys}<$900~K). 
The quasar 4C~+69.21 was not detected at all frequencies probably because of intrinsically low brightness.
Four targets -- 3C~111, 3C~120, BL~Lac, and DA~55 -- showed significant linear polarization at 22 GHz (with degrees of polarization being a few percent).
For highly polarized sources, BL Lac and DA~55, polarization could be detected at 22 and 43 GHz. 
An overview over our results is presented in Table \ref{tab:img_params}.

\subsection{3C 111}

\begin{figure}[t!]
  \centering
  \includegraphics[height=80mm,angle=-90]{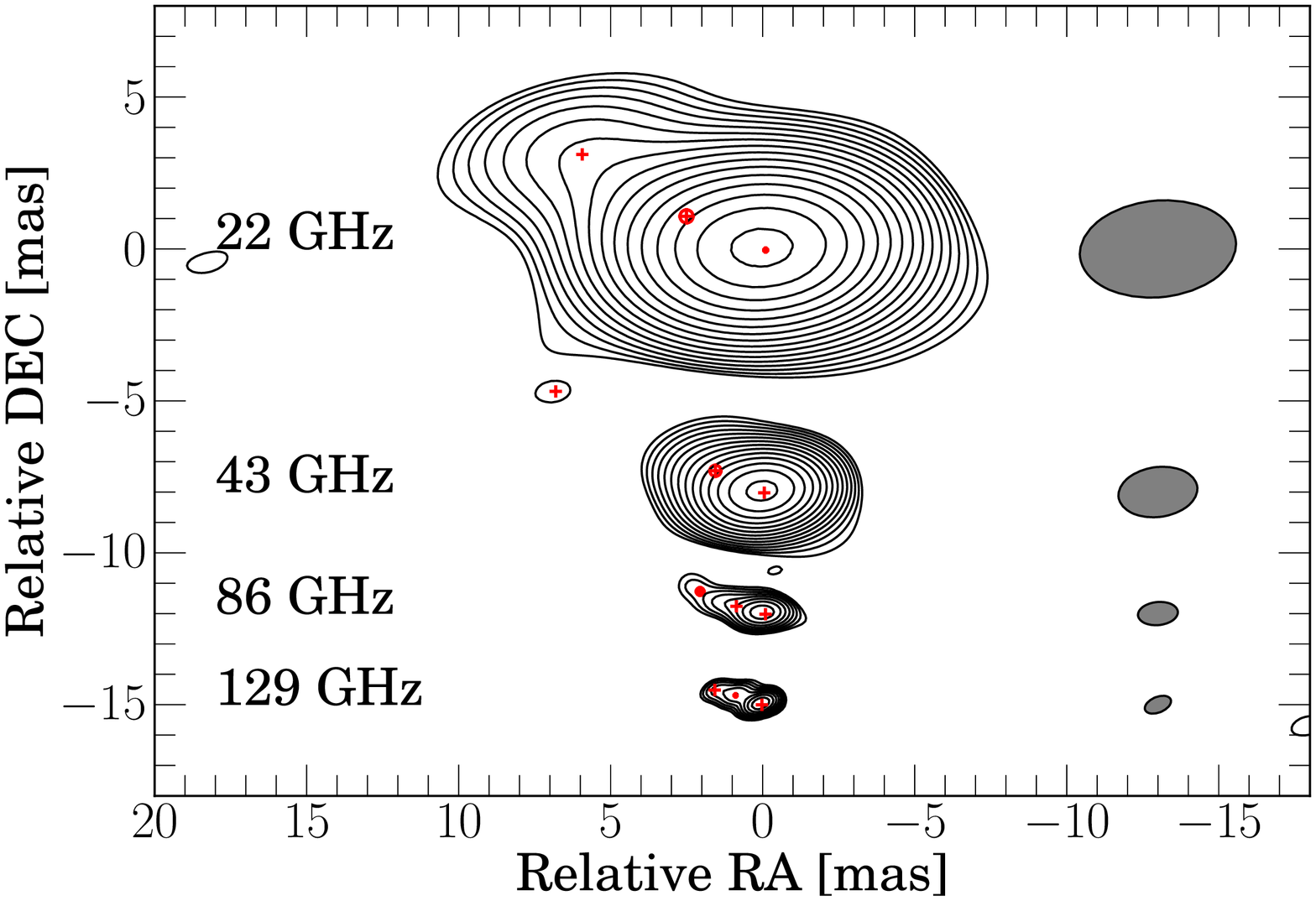}
  \includegraphics[height=45mm,angle=-90]{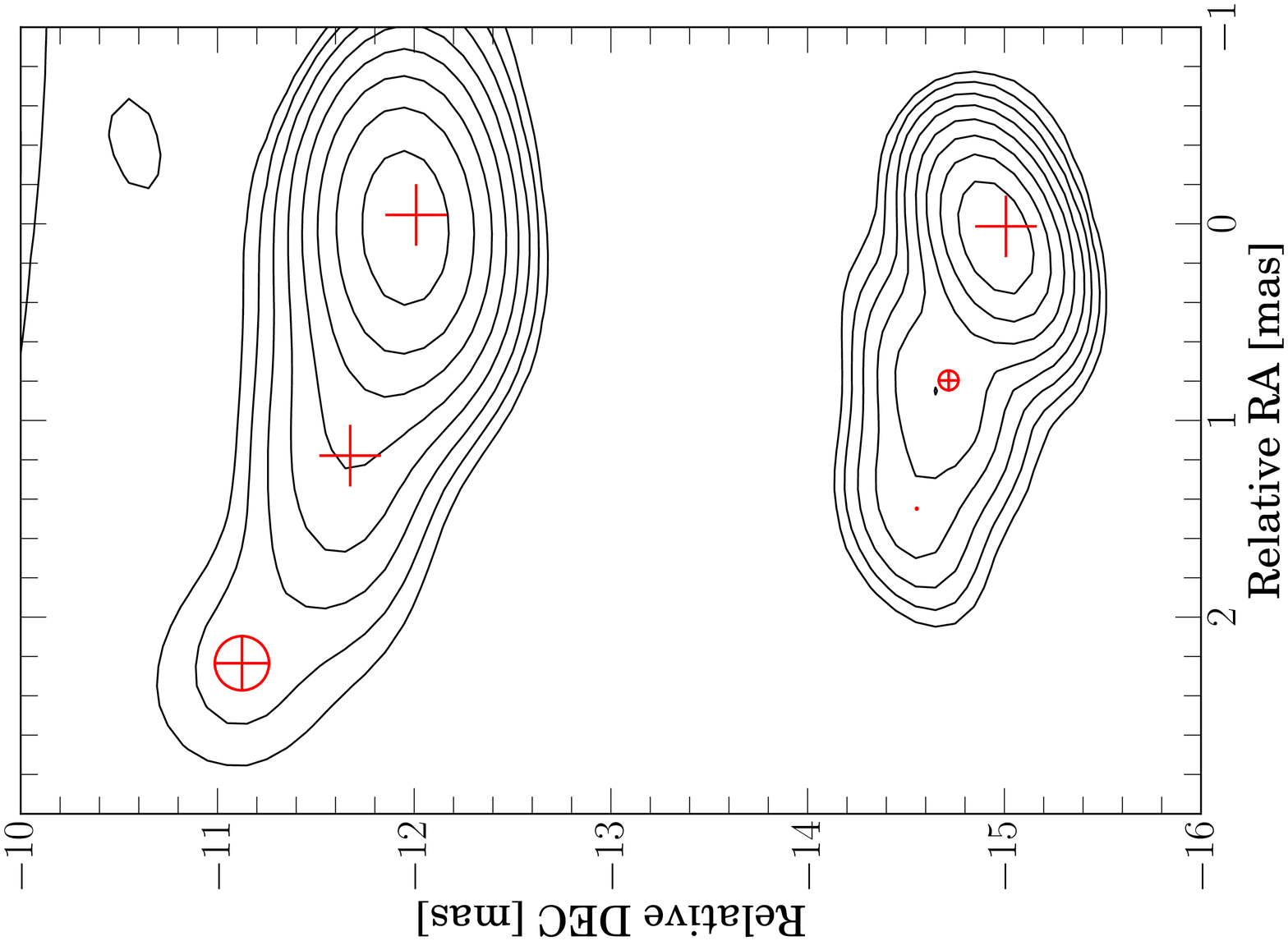}
  \caption{Stokes I maps of 3C 111.
	 Gray ellipses indicate the restoring beam size, encircled crosses represent circular Gaussian model components.
	 Model components smaller than 1\,$\mu$arcsec are indicated by crosses without circles just for a better visualization.
	 Contours increase by powers of $\sqrt{2}$.
	 \textit{Top}: Source maps for all four observing frequencies.
	 From 22 to 129 GHz, the lowest contour levels are 0.5, 1.0, 7.0, and 6.5 \% of the map peak, respectively.
	 \textit{Bottom}: Zoom onto the maps at 86 and 129 GHz.
  }
  \label{fig:3c111_imaps}
\end{figure}

\begin{figure}[t!]
  \centering
  \includegraphics[height=80mm,angle=-90]{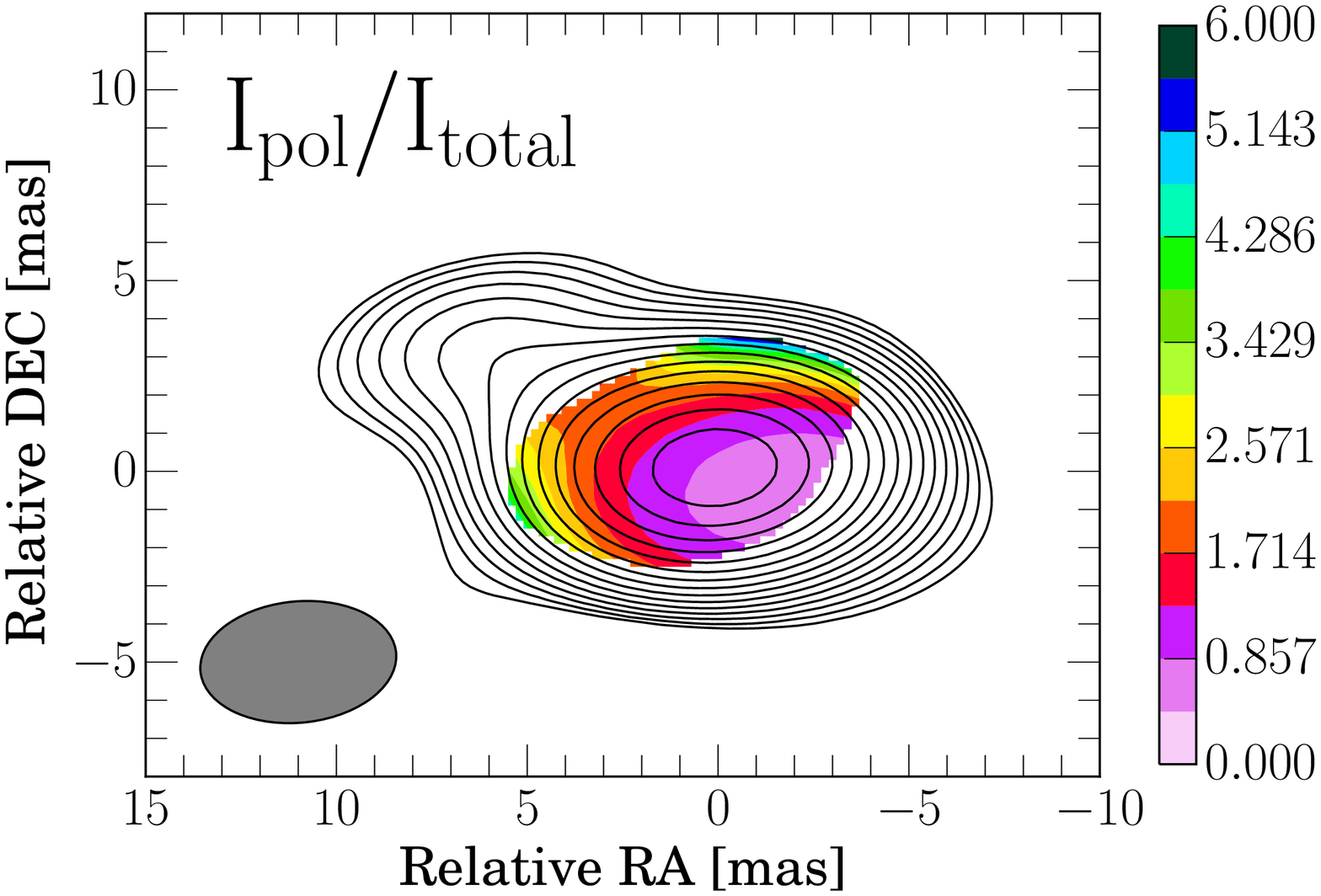}
  \includegraphics[height=80mm,angle=-90]{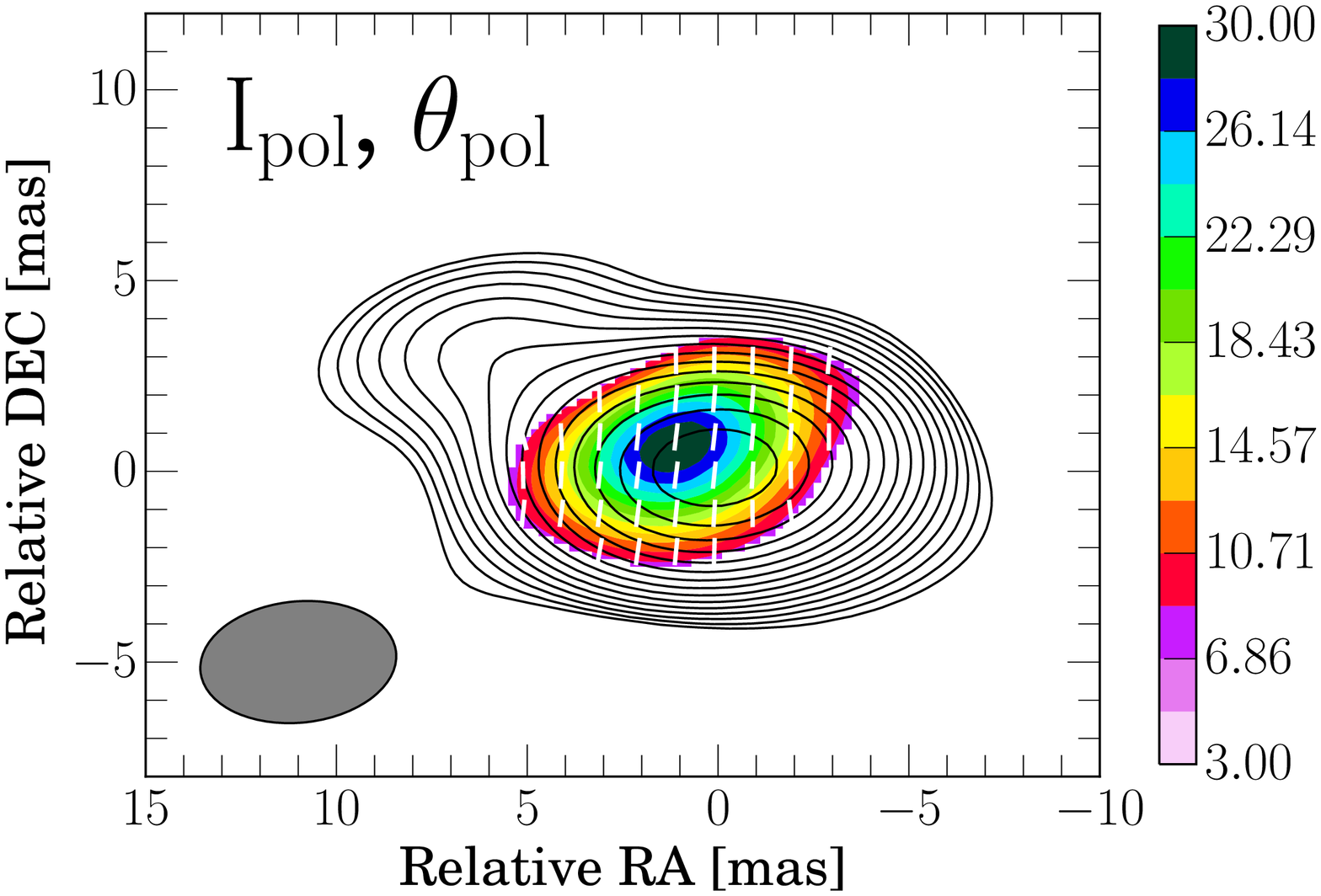}
  \caption{Polarization maps of 3C 111 at 22 GHz.
	 Black contours and gray ellipses are the same as in Figure \ref{fig:3c111_imaps}.
	 \textit{Top}: Fractional polarization map; the color bar is in units of per cent.
	 \textit{Bottom}: Polarized intensities (color bar, in mJy/beam) and relative (uncalibrated) polarization vector orientations (white bars). 
	 For the polarization, we clipped out pixels with $p_{i} < 10 \rm ~mJy/beam$.
  }
  \label{fig:3c111_polmaps}
\end{figure}

\begin{figure}[t!]
  \centering
  \includegraphics[height=80mm,angle=-90]{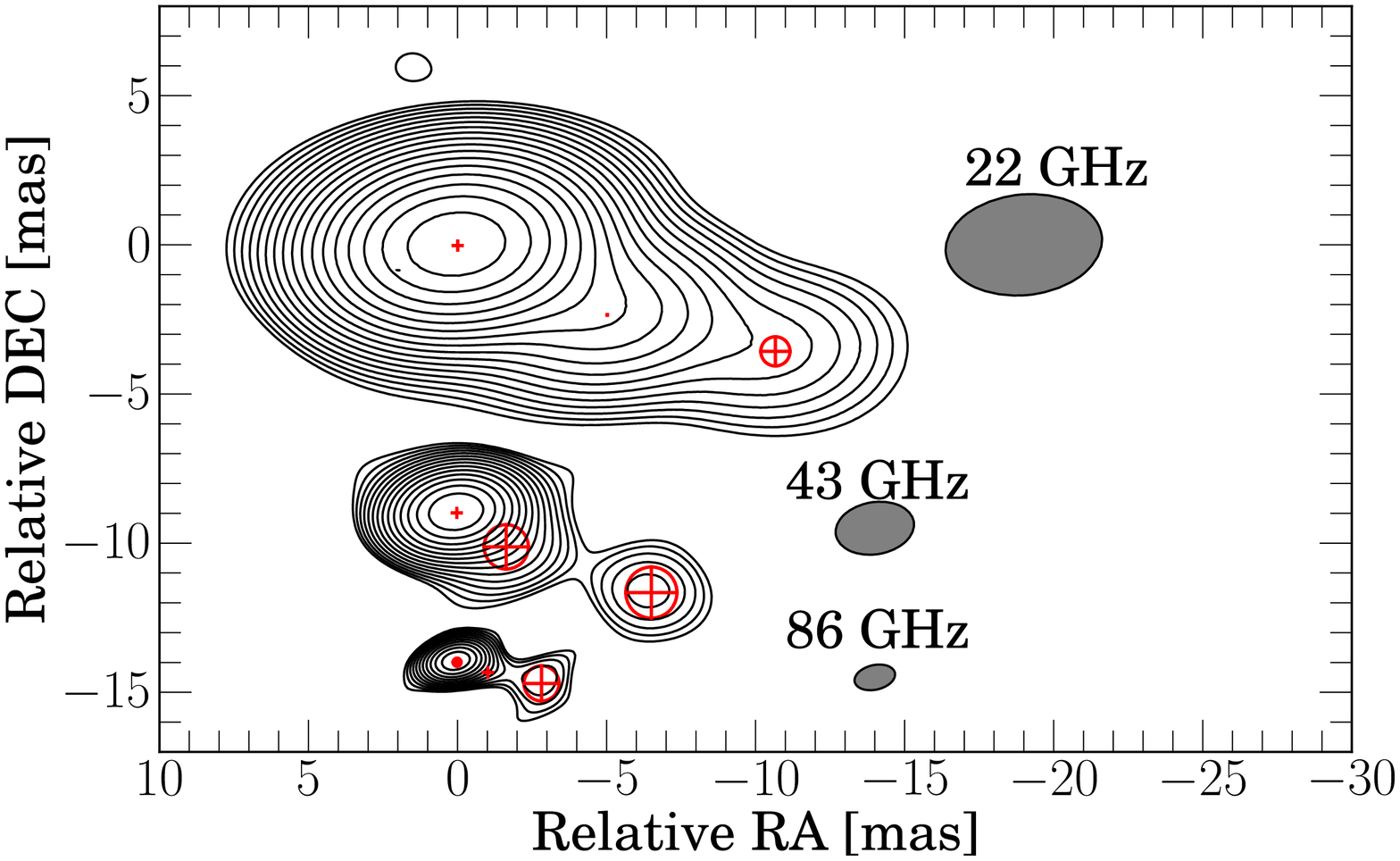} \\[8pt]  
  \includegraphics[width=60mm,angle=-90]{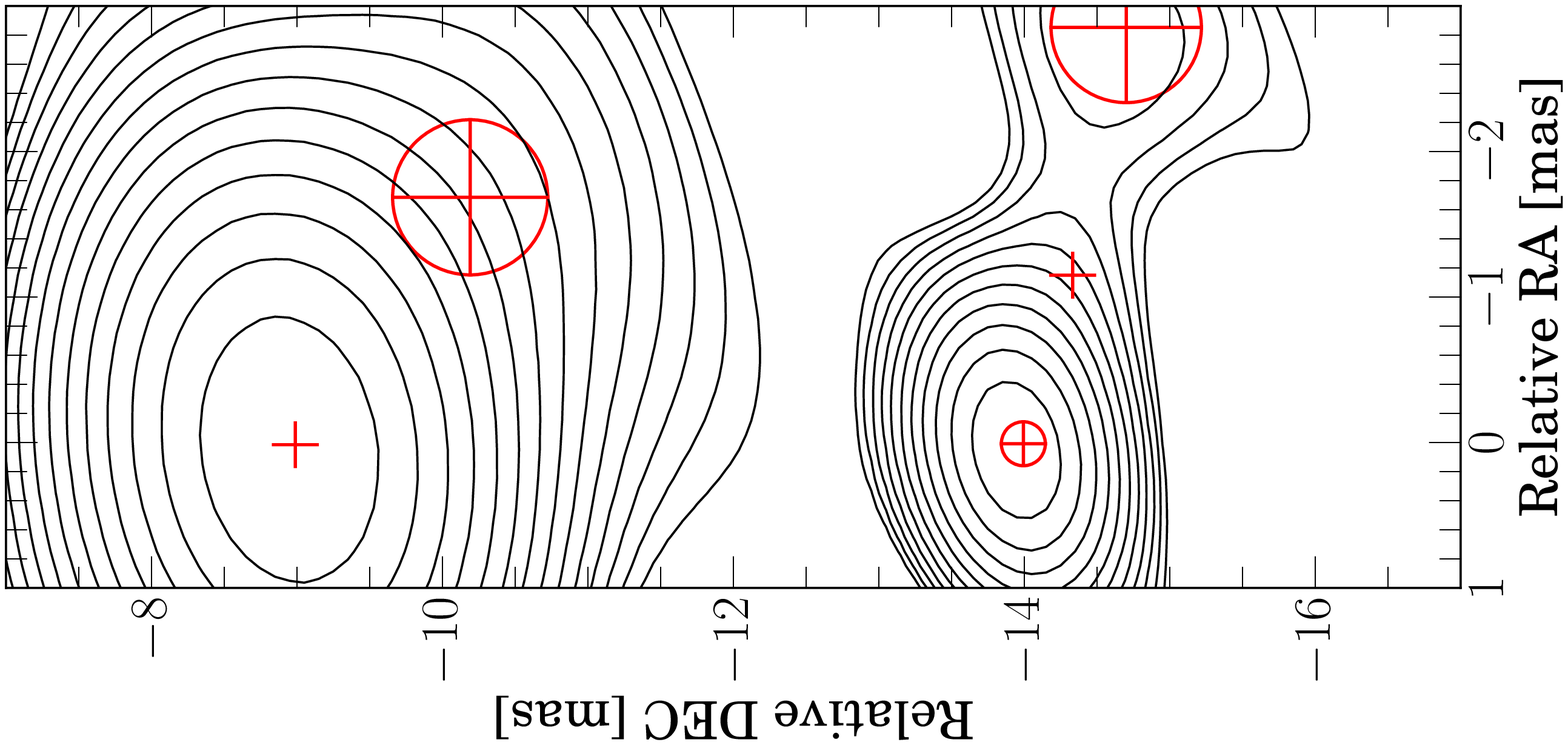}
  \caption{Stokes I maps of 3C 120.
	 \textit{Top}: Maps for three observing frequencies.
	 From 22 to 86 GHz, the lowest contour levels are 0.3, 0.8, and 2.3 \% of the map peak, respectively.
	 \textit{Bottom}: Zoom onto the nuclear regions of the maps at 43 and 86 GHz.
  }
  \label{fig:3c120_imaps}
\end{figure}

\begin{figure}[t!]
  \centering
  \includegraphics[height=80mm,angle=-90]{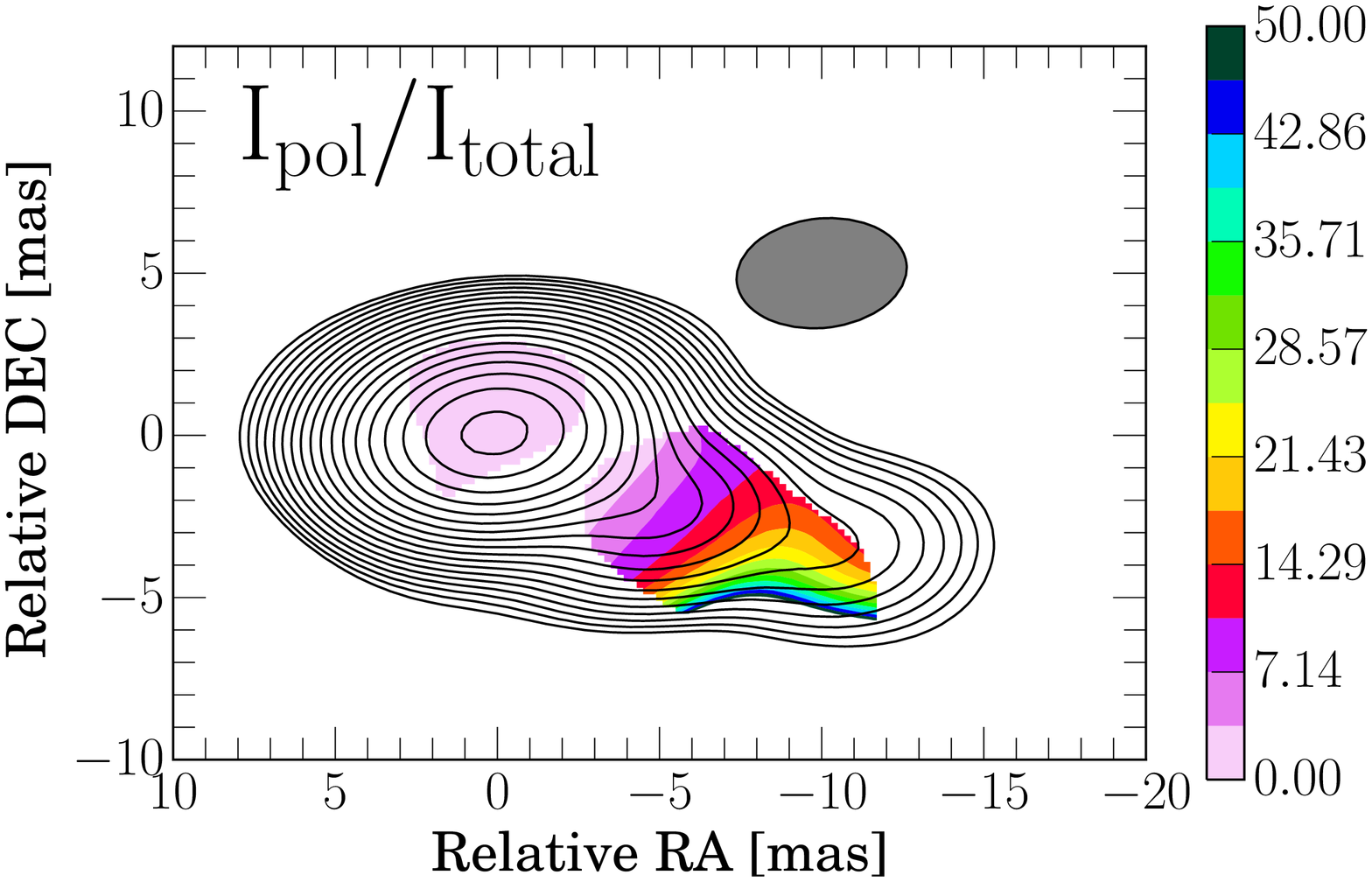}
  \includegraphics[height=80mm,angle=-90]{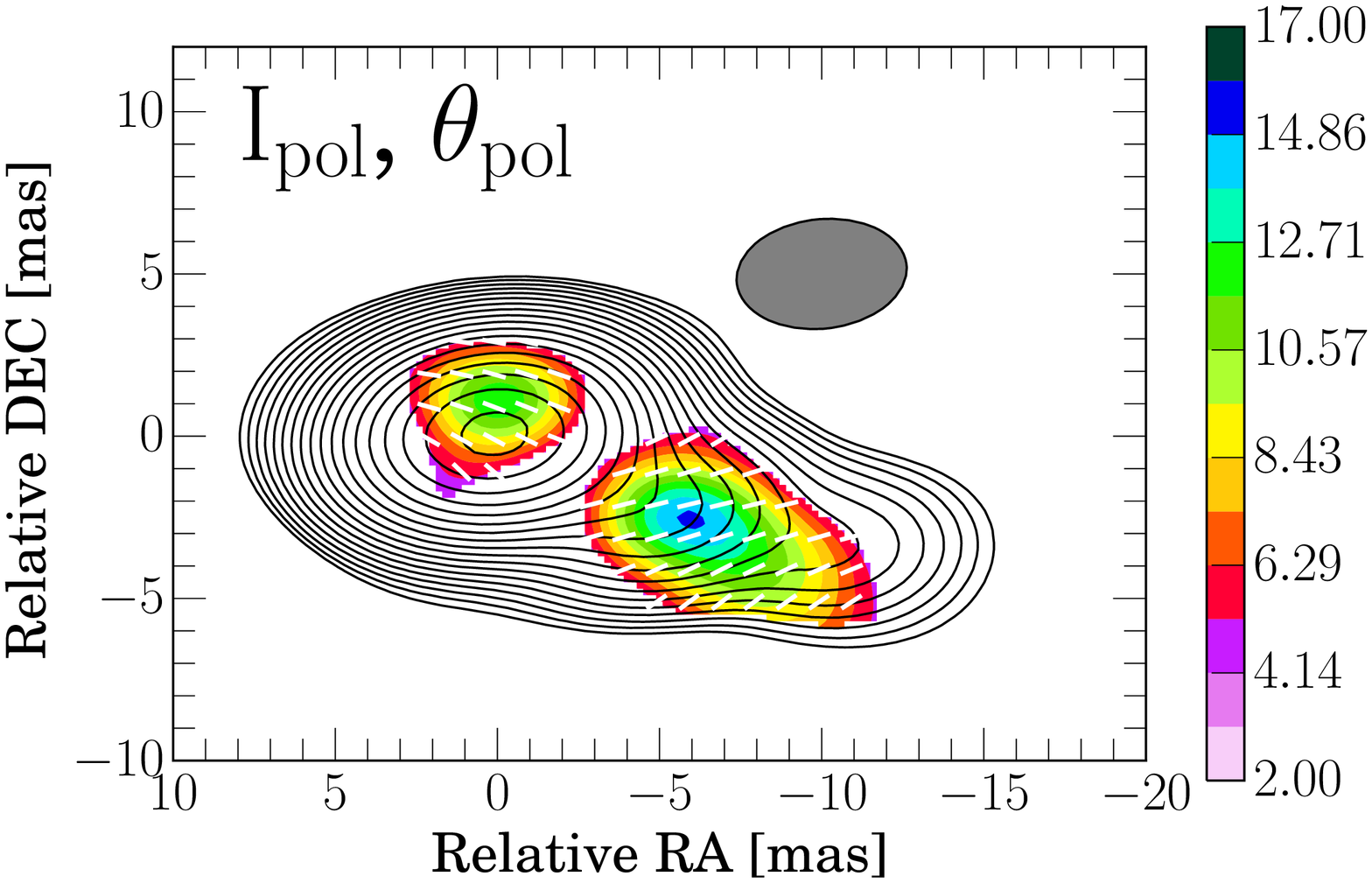}
  \caption{Same as Figure~\ref{fig:3c111_polmaps}, for 3C~120. For the polarization, pixels with $p_{i} < 4.8 ~ \rm mJy/beam$ are clipped out.}
  \label{fig:3c120_polmaps}
\end{figure}

We obtained total intensity maps of the radio galaxy 3C~111 at all four observing frequencies (Figure \ref{fig:3c111_imaps}).
3C~111 consists of three components in all maps.
For two frequency pairs (22/43 and 86/129 GHz) we were able to cross-identify components and obtained spectral information.

Figure \ref{fig:3c111_polmaps} displays the polarization structure of the source.
The level of polarization is on the order of a few per cent.
Interestingly, the peak of the polarized intensity is offset from the total intensity core. This may be caused by larger depolarization in the core due to higher opacity or by an unresolved strongly polarized jet component.

\subsection{3C 120}

3C 120 is another famous radio galaxy in our sample.
We were able to produce total intensity maps at 22, 43, and 86 GHz (Figure \ref{fig:3c120_imaps}).
For this source the cross-identification of jet components is not straightforward since the
22 GHz structure is more extended than the rather compact 43 GHz structure.
The extended emission features visible at 43 GHz may have propagated outward in the 86 GHz map 
as there is a time gap of four months between the two observations and the maximum apparent velocity of the moving features in 3C~120 is $\sim$3\,mas\,year$^{-1}$ \citep{lister13}.

Figure \ref{fig:3c120_polmaps} shows the polarized structure of 3C~120 at 22 GHz. 
The degree of polarization is a few percent. 
The peak of the polarized emission is offset from the total intensity core by $\sim$7\,mas.
Interestingly, the location of the polarization maximum is coincident with the location of the most distant jet component in the 43 GHz total intensity map.
We find different orientations of polarization for the nuclear region and the extended structure. This indicates a complex magnetic field and matter configuration in the outflow.
We note that pixels with low S/N in the total intensity and strong side lobe patterns in the linear polarization maps may have resulted in substantially high fractional polarization at the jet edges.

\subsection{3C 84}

3C 84 is a bright radio galaxy with complex structure marked by low polarization (e.g. \citealt{taylor06, trippe12, plambeck14}).
In this study we actually used 3C 84 as instrumental polarization calibrator as it is (effectively) unpolarized at frequencies $\lesssim$129~GHz; we provide only total intensity images of this source.

\begin{figure}[t!]
  \centering
  \includegraphics[height=75mm,angle=-90]{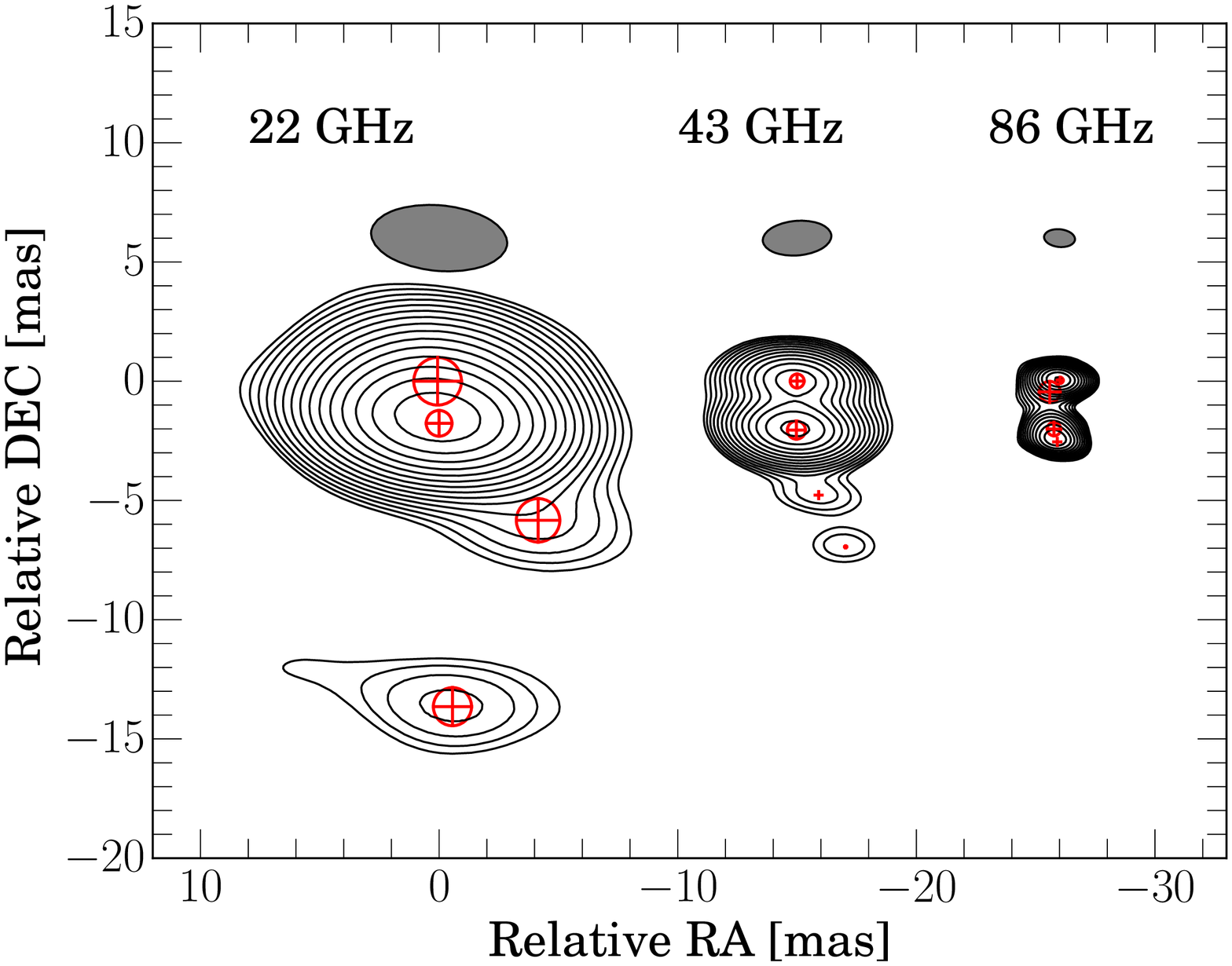}\\[3pt]
  \includegraphics[height=80mm,angle=-90,trim=0 0 0 0, clip=true]{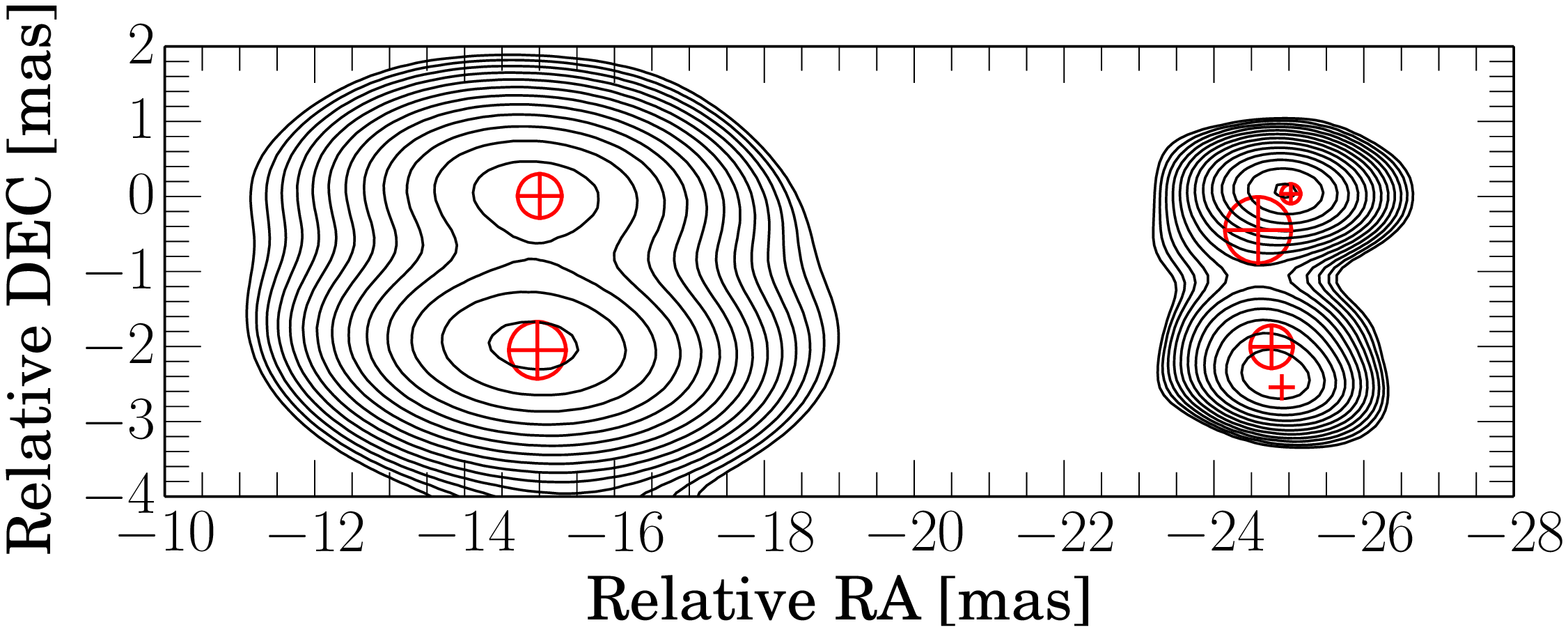}
  \caption{Stokes I maps of 3C 84.
	 \textit{Top}: Maps for three observing frequencies.
	 From 22 to 86 GHz, lowest contour levels are 0.6, 1.0, and 1.5\% of the map peak, respectively.
	 \textit{Bottom}: Zoom onto the nuclear regions of the 43 and 86 GHz maps.
  }
  \label{fig:3c84_imaps}
\end{figure}

Figure \ref{fig:3c84_imaps} displays 22, 43, and 86 GHz images of the source.
Due to technical problems with the 129 GHz receiver we did not obtain a map at 129 GHz.
Our observations show the southern jet albeit not the well-known northern counter-jet.
At 22 GHz, the source structure is only marginally resolved.
At 86 GHz, the inner region of the source is clearly resolved into two bright components.
The bright southern component ejected during the radio burst in 2005 (cf. \citealt{nagai10}) is still detectable at 86 GHz.

\subsection{4C +01.28}

\begin{figure}[t!]
  \centering
  \includegraphics[height=80mm,angle=-90]{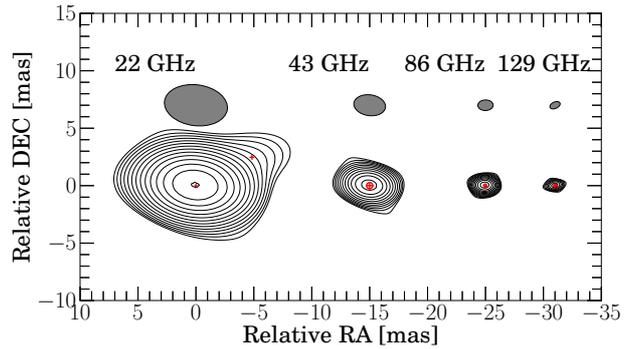}
  \caption{Stokes I maps of 4C +01.28 for four observing frequencies.
	 From 22 to 129 GHz, lowest contour levels are 1.1, 2.7, 2.5, and 10.0 \% of the map peak, respectively.
  }
  \label{fig:4c0128_imaps}
\end{figure}

The quasar 4C +01.28 has the highest redshift among all sources in our sample ($z=0.888$).
Accordingly, the source appears compact at all frequencies (cf. Figure~\ref{fig:4c0128_imaps}).
We did not detect significant polarization.
A faint extended jet in northwestern direction was detected at 22~GHz.

\subsection{BL Lac}

\begin{figure}[t!]
  \centering
  \includegraphics[height=80mm,angle=-90]{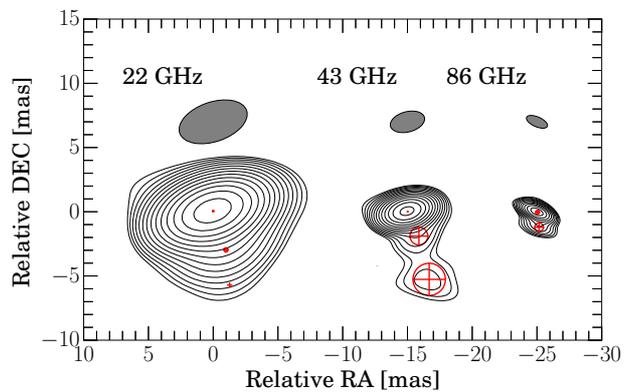}
  \caption{Stokes I maps of BL Lac for three observing frequencies.
	 From 22 to 86 GHz, lowest contour levels are 0.9, 1.4, and 3.5 \% of the map peak, respectively.
  }
  \label{fig:bllac_imaps}
\end{figure}

The radio source BL Lacertae is (among other characteristic features) characterized by a high degree of linear polarization.
We obtained polarization maps at 22 and 43 GHz.
Figure \ref{fig:bllac_imaps} shows the total intensity maps for 22, 43, and 86~GHz; at 129 GHz, no fringe solutions for the LL and RR visibilities were found. 
The 22 and 43 GHz maps show three Gaussian components each for which cross-identification was straightforward.
For the model components found in the 86 GHz map, no counterparts at 129~GHz are available due to lack of 129 GHz data.

Figure \ref{fig:bllac_polmaps} displays distributions of fractional polarization and polarized intensity.
In spite of our failure to derive absolute EVPA and thus rotation measure maps,
the polarization fraction $m_{L}$ and relative EVPA obtained at several frequencies already illuminate some of the source physics.
A weak $m_L$ gradient is present in the direction of the outflow in the 22 GHz map. 
The level of polarization at the core is similar ($\sim$3\%) at both frequencies. The extended jet component in the 43 GHz map shows very substantial polarization of almost 40\%. We note that the high degree of polarization at the edge of the core shown in the 43-GHz map is located in a region of weak total intensities and strong side lobe patterns in the polarization map.

\subsection{DA 55}

\begin{figure}[t!]
  \centering
  \includegraphics[height=80mm,angle=-90]{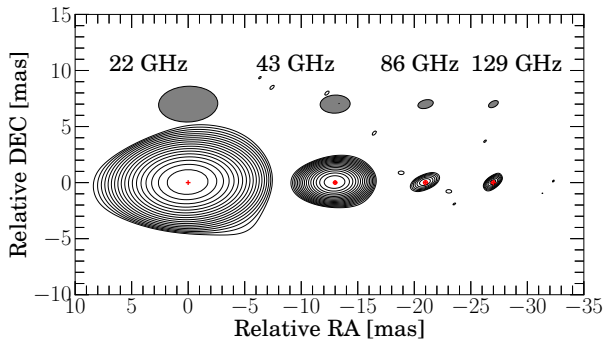}
  \caption{Stokes I maps of DA 55 for four observing frequencies.
	 From 22 to 129 GHz, lowest contour levels are 0.2, 0.4, 13.0, and 10.0 \% of the map peak, respectively.
  }
  \label{fig:da55_imaps}
\end{figure}

DA~55 is a compact and highly polarized quasar at redshift $z=0.859$. It has a compact structure and is dominated by the core (Figure~\ref{fig:da55_imaps}).
Although the structure is only marginally resolved, we detected and mapped significant linear polarization at both 22 and 43~GHz.
Figure~\ref{fig:da55_polmaps} shows the spatial distribution of $m_{L}$ and polarized intensity.
At both frequencies, DA~55 shows $m_{L}\sim4$\% and well-ordered polarization angles across the source.

\begin{table}[t!]
\caption{Summary of the $T_{b}$ distributions.\label{tab:tb}}
\centering
\begin{tabular}{lcc}
\toprule
Components$^{\rm a}$ & $\log T_{b} \pm \sigma_{T_{b}}$ $^{\rm b}$ [K] & \#$^{\rm c}$ \\ 
\midrule
\textbf{Resolved} && \\ 
All  & $8.76 \pm 0.98$ & 6 \\
Core components & -- & 0 \\
Jet components & $8.76 \pm 0.98$ & 6 \\
\addlinespace 
\textbf{Unresolved} && \\ 
All  & $> 8.89$ & 44 \\
Core components & $> 9.86$ & 21 \\
Jet components & $> 8.01$ & 23 \\
\bottomrule
\end{tabular}
\tabnote{
Distributions for spatially resolved and unresolved components, at all frequencies.
(a): Source regions analyzed; 
(b): $T_{b}$ values with standard deviations (resolved components), lower limits on $T_{b}$ (unresolved components);
(c): Number of Gaussian components.
}
\end{table}

\section{Discussion\label{sec:discussions}}

In this section we discuss brightness temperature distributions, possible correlations between physical parameters, and potential physical interpretations of observed relations.

\subsection{Brightness Temperature Distributions}

We calculated $T_{\rm b,app}$ distributions for all emission features (Gaussian model components) at all observing frequencies.
Figure~\ref{fig:Tb_info} shows histograms of $T_{\rm b,app}$ measured for both cores and extended jet components at the all frequency range
and Table~\ref{tab:tb} summarizes the results.
The average value of our \textit{measured} $T_{b}$ is $10^{8.76 \pm 0.98}$~K for the jets and
the averaged \textit{lower limits} are $10^{9.86}$~K and $10^{8.01}$~K for core and jet components, respectively (all averages are averages of logarithms). 
In multiple cases we found that the Gaussian model components were smaller than $d_{\rm lim}$; this is probably characteristic of KVN data as we briefly mentioned in the Section \ref{sec:analysis}. 
Therefore we obtained only lower limits for $T_{\rm b,app}$ via Equation~(\ref{eq:tb}).

The values we observe in our $T_{\rm b,app}$ distributions are difficult to compare to other reference values,
for example the brightness temperature for energy equipartition of a synchrotron-emitting plasma, 
$T_{\rm b}\sim5\times10^{10}$\,K \citep{readhead94}.
This is due to small number statistics as well as our rather limited size measurement capability, which leads to lower limits only. 
A further source of uncertainty is Doppler boosting; depending on the relativistic bulk speed of an emitter, the intrinsic brightness temperature
$T_{\rm b,int}$ will be lower than $T_{\rm b,app}$ by the Doppler factor $\delta=1/\Gamma(1-\beta_{\Gamma}\cos \theta)$,
where $\Gamma$ is the bulk Lorentz factor of the propagating material, $\beta_{\Gamma}$ is the associated speed in units of the speed of light, and $\theta$ is the angle relative to the line of sight.
If we take $\delta \sim 10$ (e.g., \citealt{jorstad05}) the intrinsic $T_{\rm b}$ of the jet components will be lower than the measured ones \
by one order of magnitude -- which implies a large uncertainty.
For this reason, long-term monitoring of the targets with future observations is necessary to complement $\delta$ information from jet kinematics.

\subsection{Correlations between Parameters}

We searched for possible correlations between spectral and polarimetric properties of the cores and jet components.
As a first step, we computed the spectral index $\alpha$.
Figure~\ref{fig:spix_info} shows $\alpha_{22-43 \rm GHz}$ and $\alpha_{86-129 \rm GHz}$ versus observed 22 and 86~GHz fluxes and (for jet components) projected distances from the cores.
The spectral indices of the cores measured between 22 and 43 GHz are essentially consistent with zero regardless of flux.
This result is reasonable since the observing frequencies of KVN are high enough to be free from synchrotron self-absorption, which
leads to inverted spectra at lower frequencies (e.g., \citealt{hovatta14}).
The jets have steeper spectra in general but do not show an obvious spectral evolution as function of projected distance (which might be due to the small sample size).
It is worthwhile to note that at high frequencies (86--129 GHz) the cores show rather steep spectra, i.e., the core fluxes at 129~GHz are substantially lower than those at 86~GHz.
Since the $uv$ radii of KVN at 86 and 129 GHz differ only by factor of 1.5, this difference is unlikely to be an effect of incomplete $uv$ coverage. (The spectral steepening of the cores will be discussed in a dedicated follow-up paper).

\begin{figure*}[t!]
  \centering
  \includegraphics[width=55mm,angle=-90]{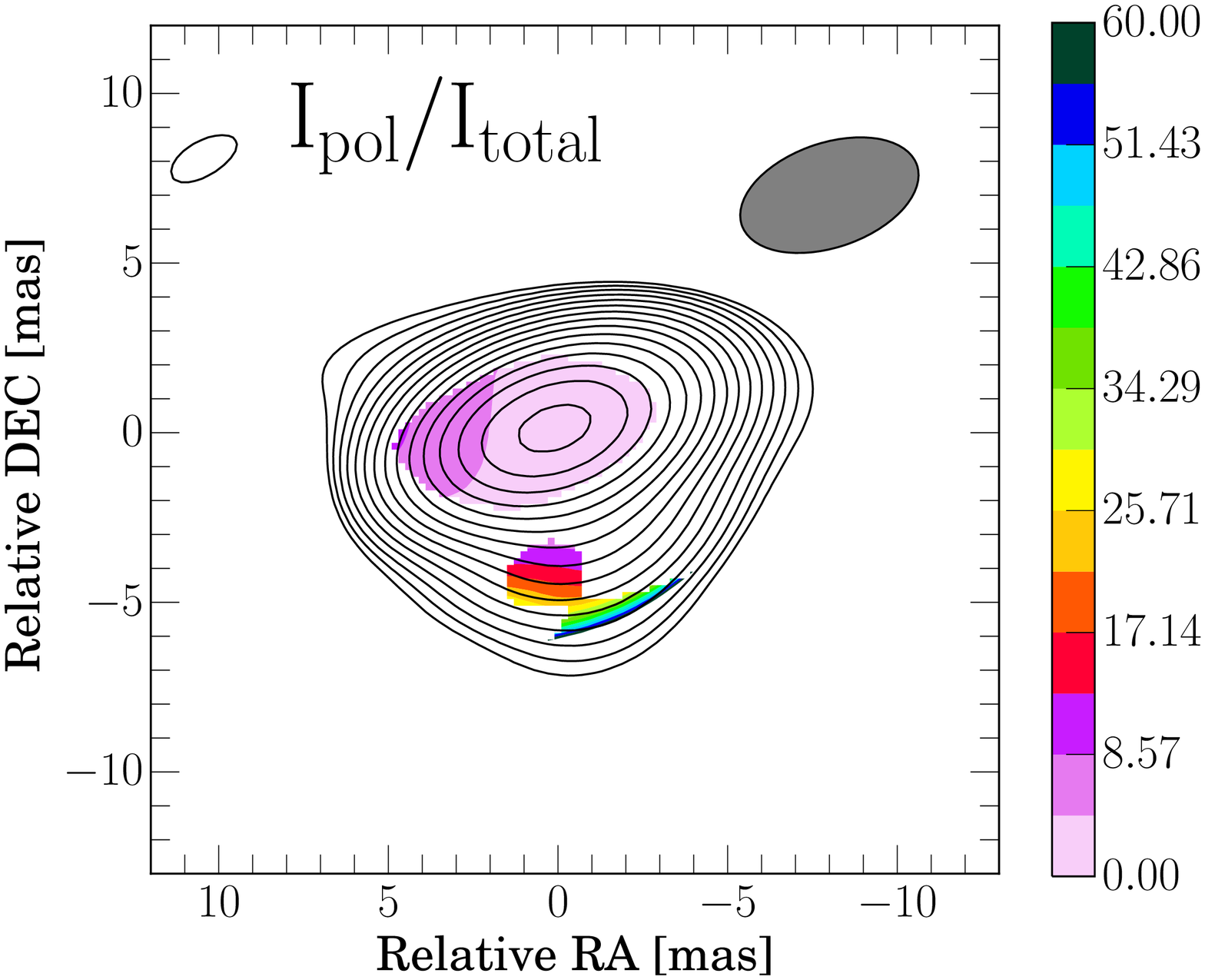} \hspace{5mm}  
  \includegraphics[width=55mm,angle=-90]{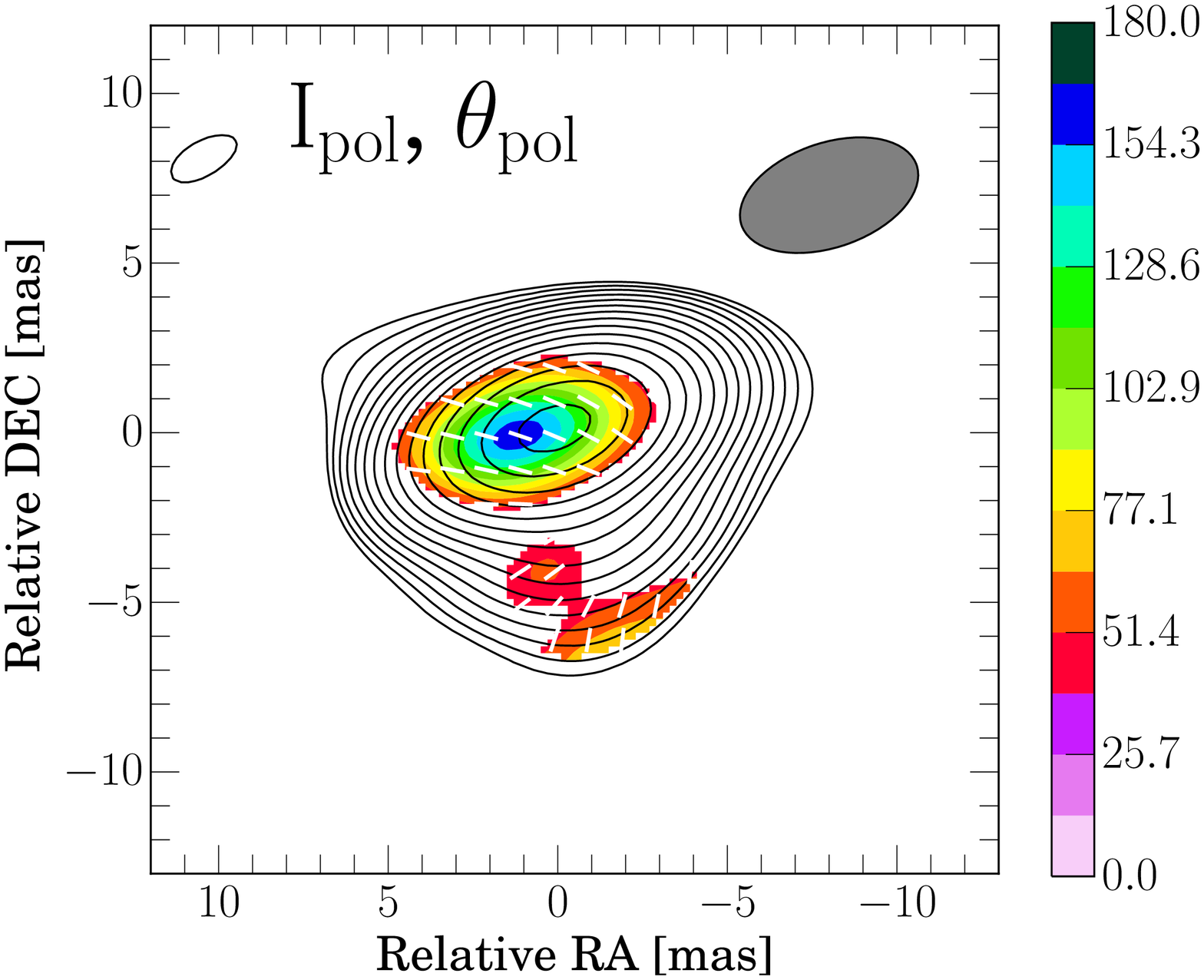} \\ 
  \includegraphics[width=55mm,angle=-90]{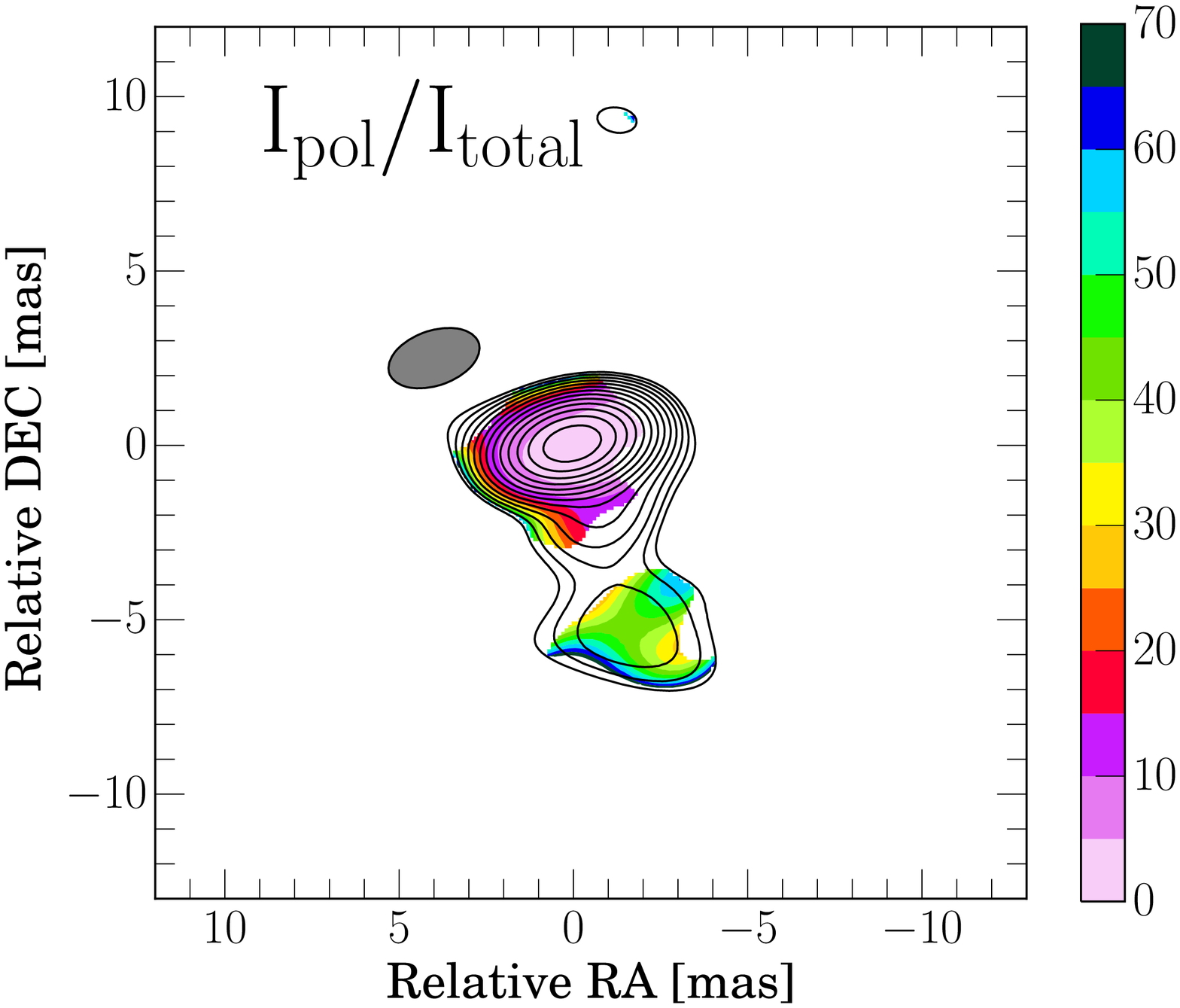} \hspace{5mm}  
  \includegraphics[width=55mm,angle=-90]{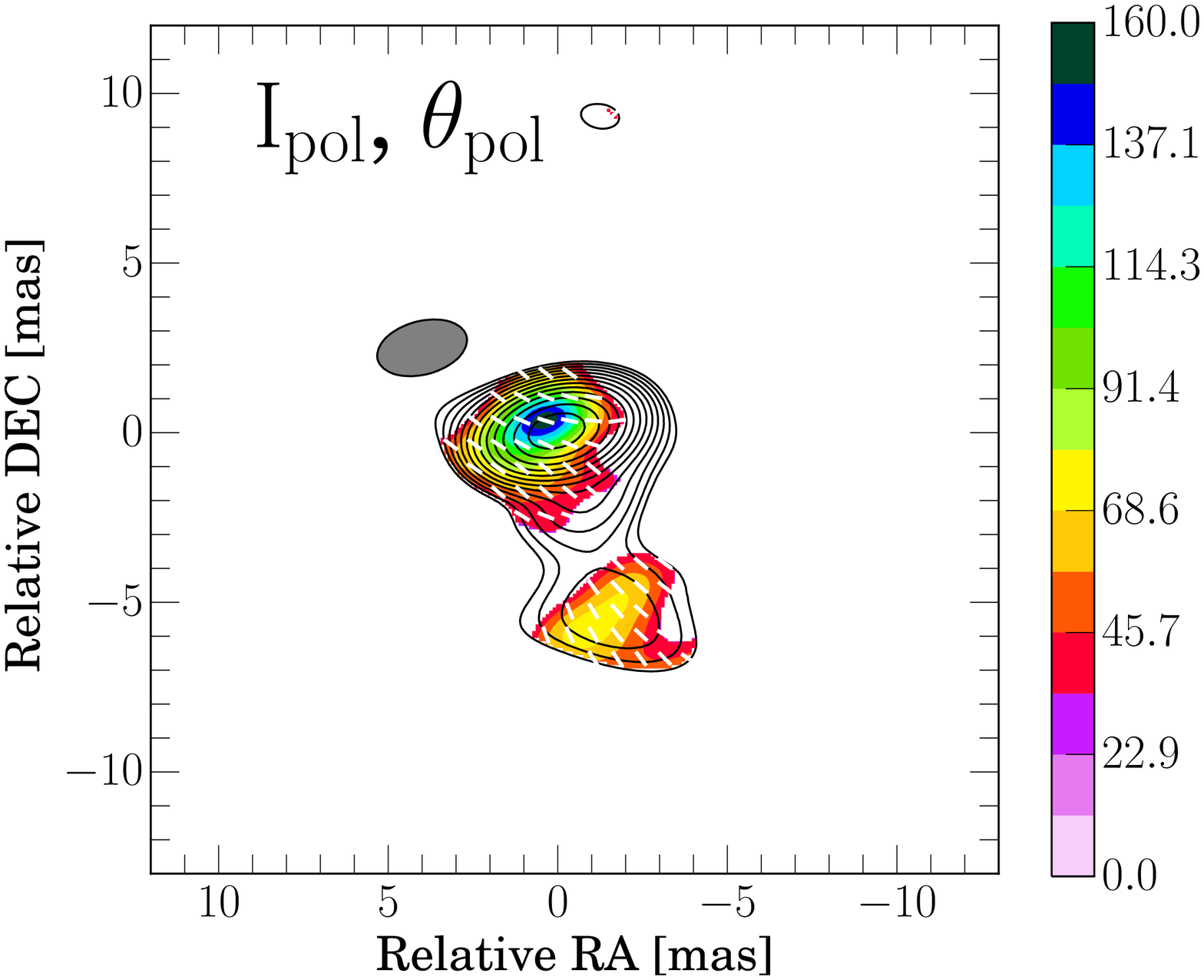}
  \caption{Polarization maps of BL Lac at 22 GHz (top panels) and 43 GHz (bottom panels).
	 Black contours and gray ellipses are the same as in Figure \ref{fig:bllac_imaps}.
	 \textit{Top left}: Fractional polarization maps at 22 GHz; the color bar is in units of per cent.
	 \textit{Top right}: Polarized intensities at 22 GHz (color bar, in mJy/beam) and relative (uncalibrated) polarization angles (white bars). 
	 \textit{Bottom left}: Fractional polarization map at 43 GHz. 
	 \textit{Bottom right}: Polarized intensities and relative polarization angles at 43 GHz.
	 For the polarization, pixels with $p_{i} <$ 45 and 32.7 $\rm~ mJy/beam$ for 22 and 43 GHz, respectively, are clipped out.
  }
  \label{fig:bllac_polmaps}
\end{figure*}

\begin{figure*}[t!]
  \centering
  \includegraphics[width=55mm,angle=-90]{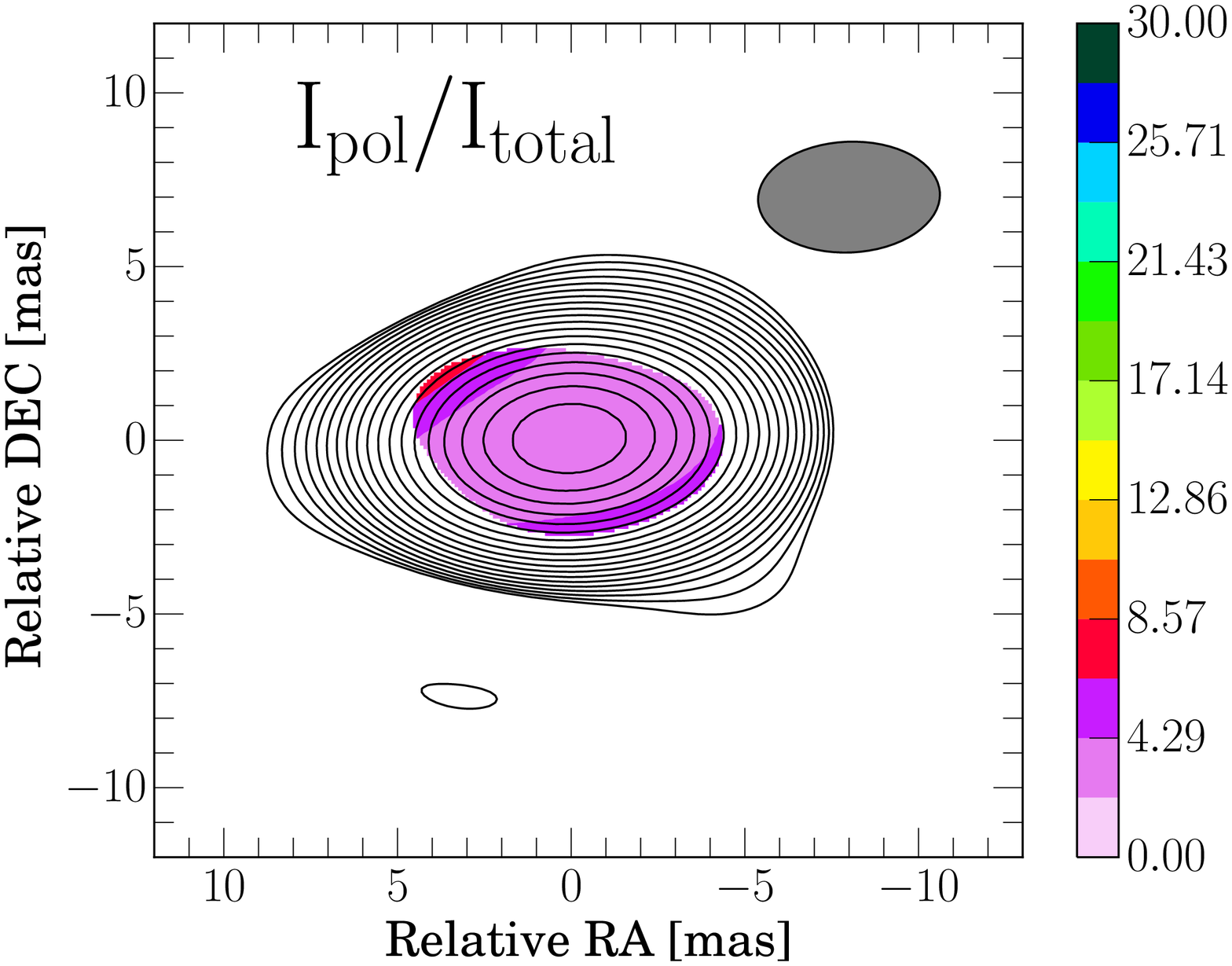} \hspace{5mm}  
  \includegraphics[width=55mm,angle=-90]{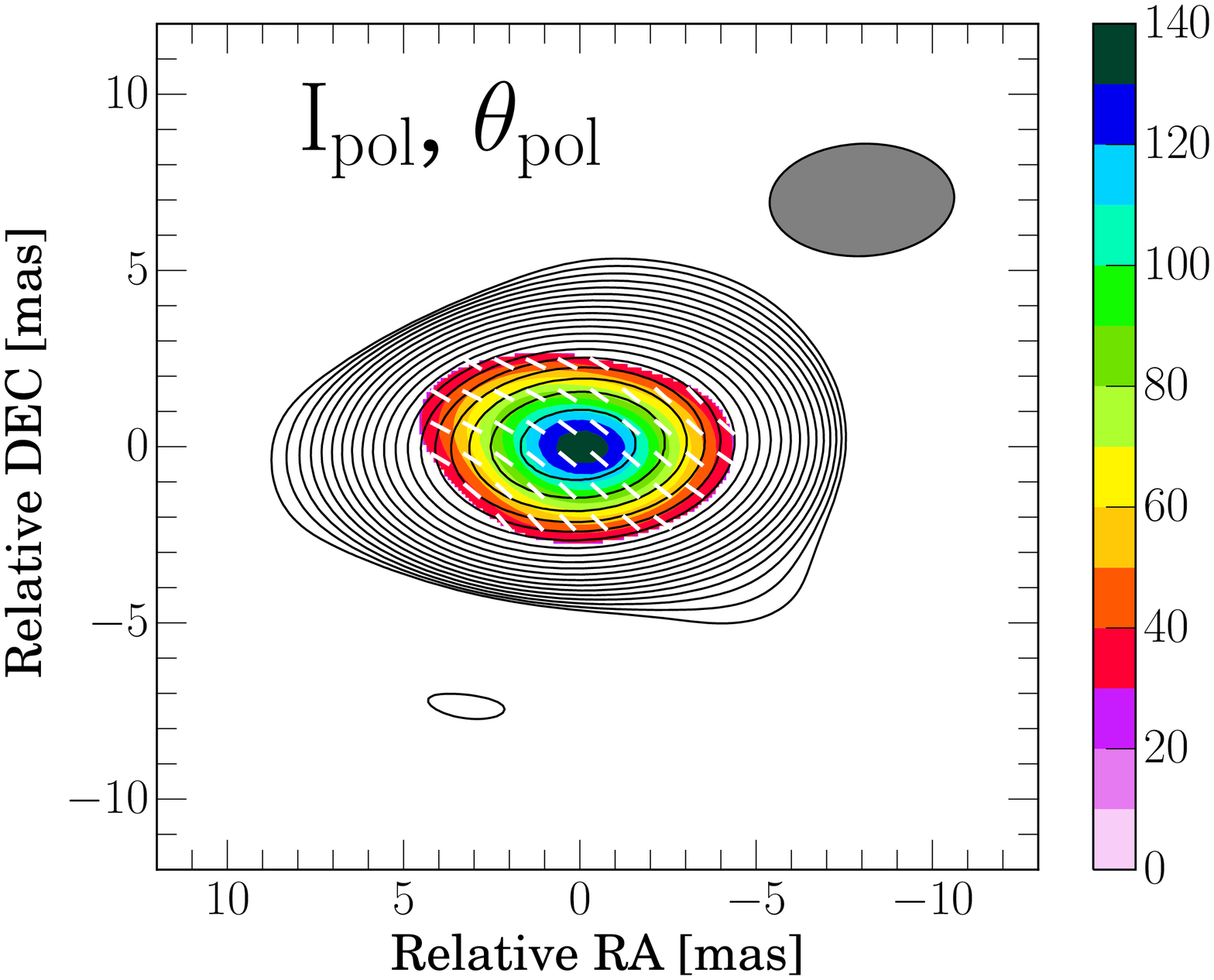} \\  
  \includegraphics[width=55mm,angle=-90]{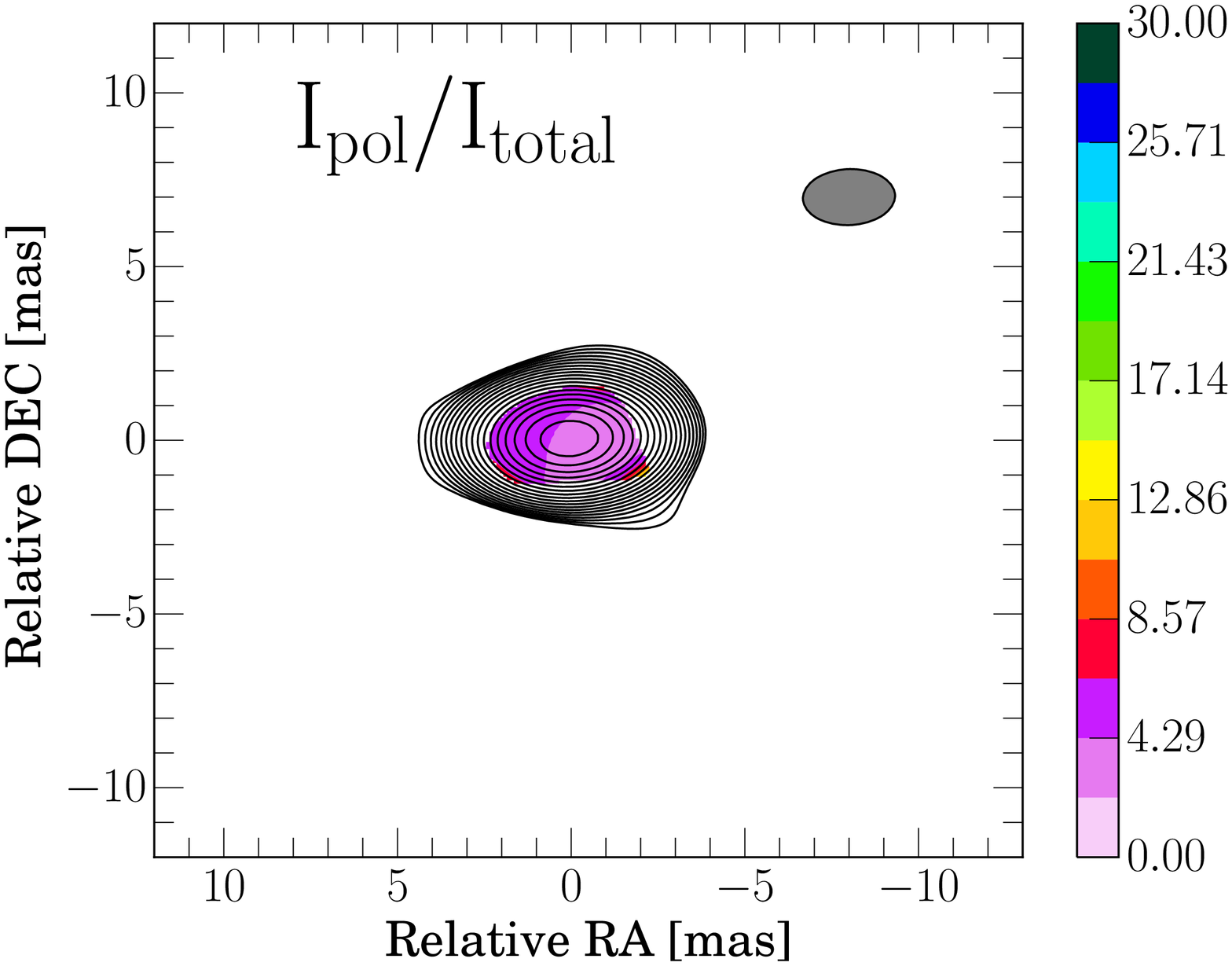} \hspace{5mm}  
  \includegraphics[width=55mm,angle=-90]{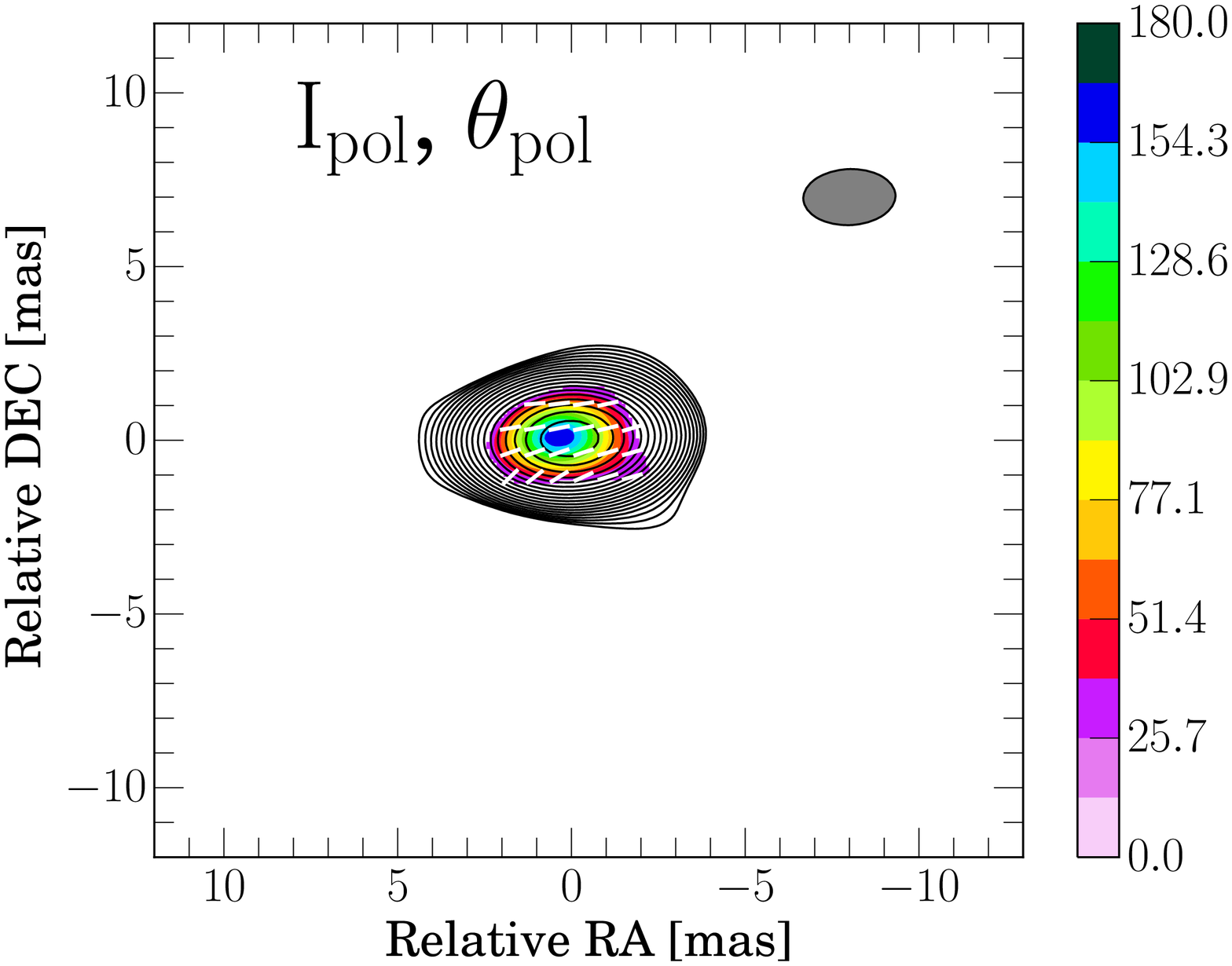}
  \caption{Same as Figure \ref{fig:bllac_polmaps}, for DA 55. 	 
  For the polarization, pixels with $p_{i} <$ 28.2 and 25.2 $\rm~ mJy/beam$ for 22 and 43 GHz, respectively, are clipped out.}
  \label{fig:da55_polmaps}
\end{figure*}

Figure~\ref{fig:pol_info} shows the degree of polarization of the cores and jet components as function of flux and (for jet components) projected distance from the cores.
The level of polarization for both jets and cores is about or less than 10\%, in agreement with the results of previous studies (e.g., \citealt{lister05, jorstad07, trippe10}).
We do not find indications for a correlation between $m_{L}$ and flux. 
The fractional polarizations of jet components tend to increase with projected distance.
Such behavior was already found by other VLBI observations (e.g., see Figure~6 of \citealt{lister05} and references therein), albeit no physical mechanism was suggested then.
The Faraday depolarization caused by a foreground medium might change the observed level of polarization
as function of the projected distance. For kpc-scale AGN with twin radio jets, \citet{garrington88} showed that the emission from jets pointing away from the observer is affected by larger depolarization than jets pointing toward the observer because of the concentration of interstellar matter in the galactic core.
For pc-scale one-sided jets this scenario implies that more distant jet components are located in regions with lower densities, and thus lower column densities, of depolarizing interstellar matter.
However, as on parsec scales the interstellar medium is dominated by dusty tori rather than spherical gas distributions \citep[e.g.,][]{walker00, kameno01, kadler04}, explaining the observed polarization pattern as a column density effect is not straightforward. Alternatively, the source-intrinsic linear polarization might increase with decreasing particle density in an expanding outflowing
jet component. The relation between the plasma opacity (spectral index) and the degree of linear polarization is discussed separately in the following.

\begin{figure}[t!]
  \centering
  \includegraphics[width=80mm]{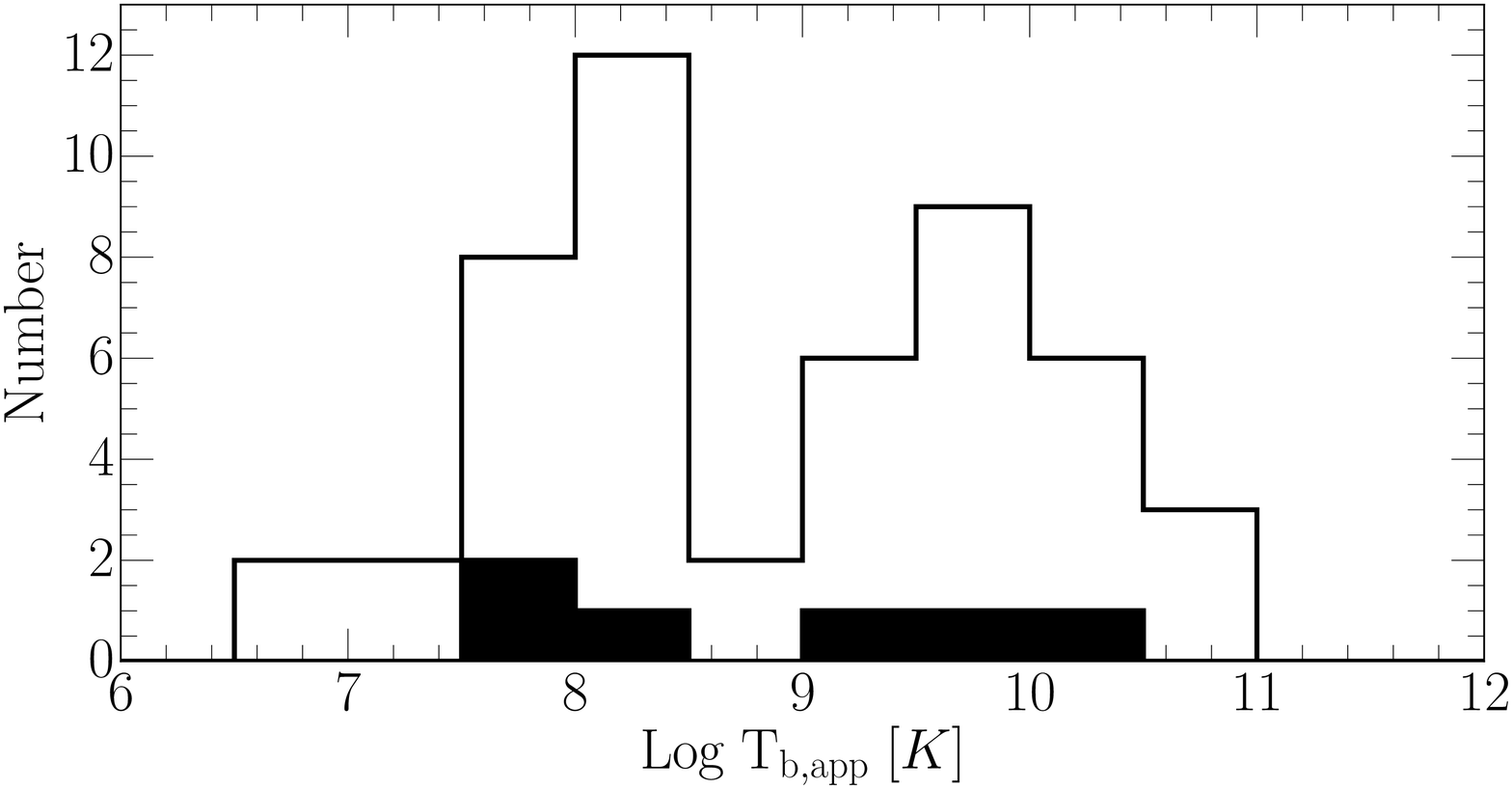}
  \includegraphics[width=80mm]{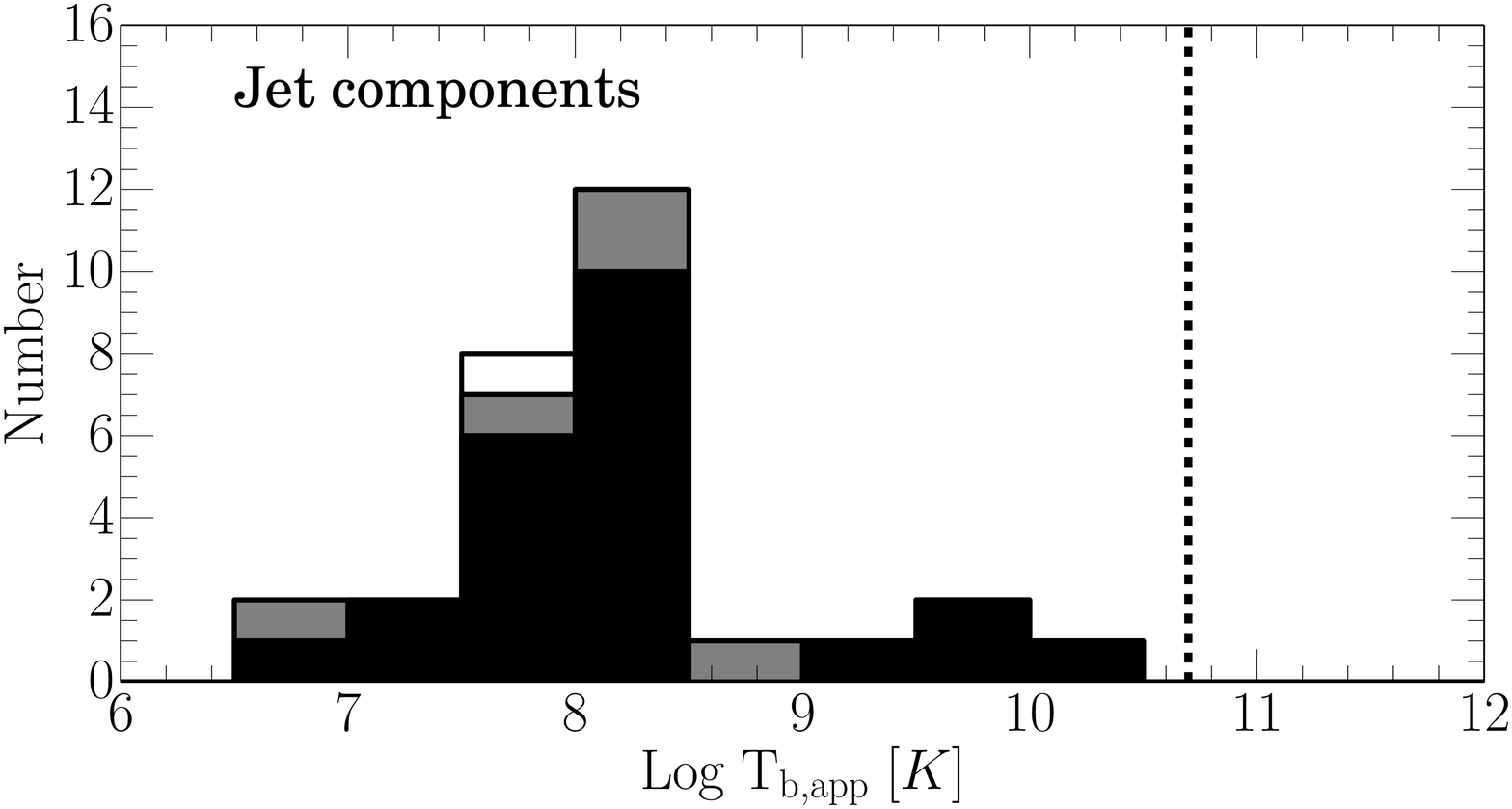}
  \includegraphics[width=80mm]{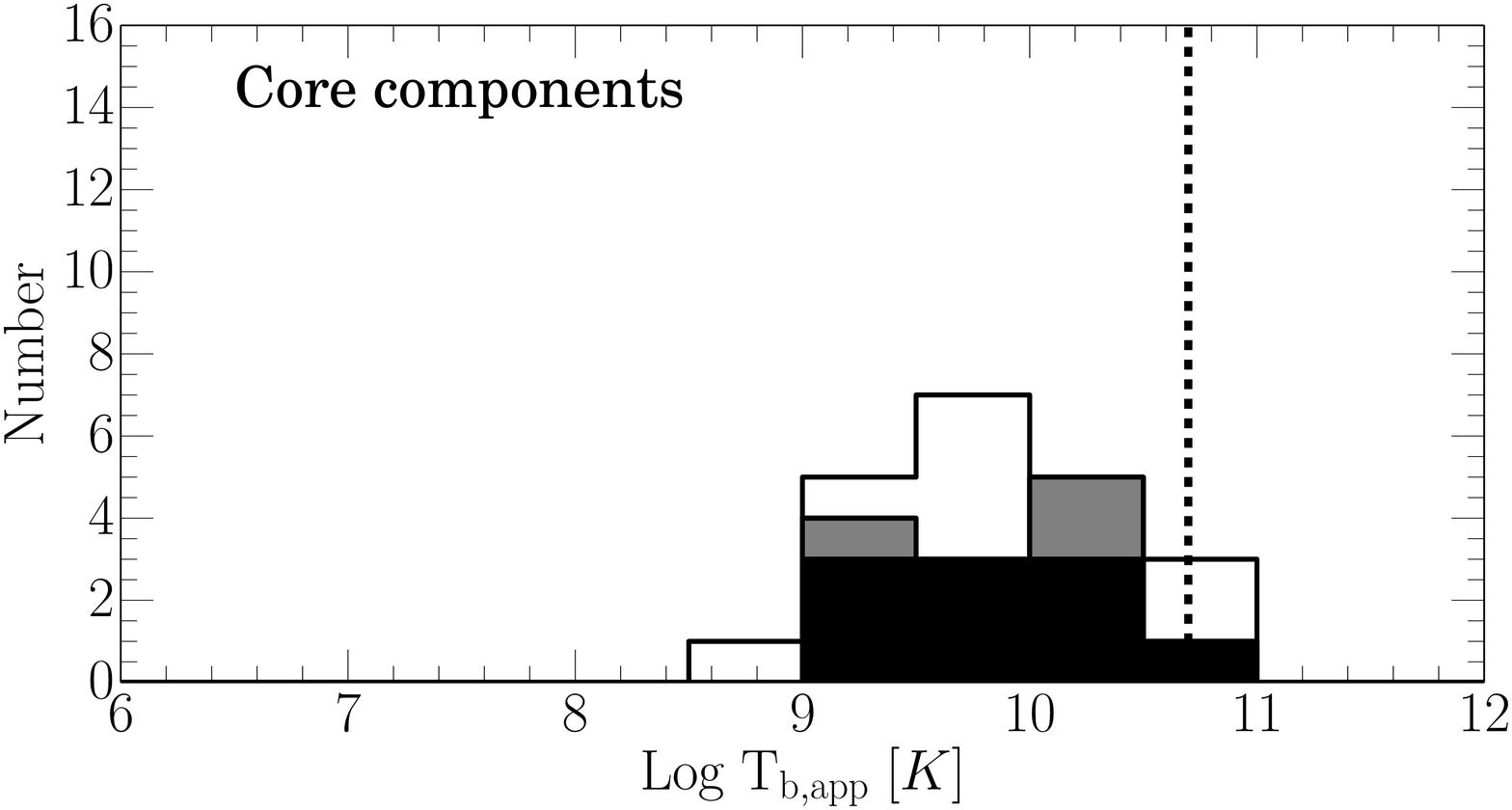}
  \caption{Brightness temperature distributions for Gaussian model components found in KVN maps.
	 \textit{Top}: Histograms of $T_{\rm b,app}$. The bin size is 0.5 in units of $\log T_{\rm b,app}$.
	 Empty and filled areas indicate lower limits (for unresolved components) and actually measured values (for resolved components), respectively.
	 \textit{Center}: The same histogram for extended jet components only, distinguishing radio galaxies (3C 111, 3C 120, and 3C 84, in black), BL Lac (in gray), and quasars (4C +01.28 and DA 55, in white).
	 The dotted line corresponds to $T_{\rm b,app}=5\times10^{10}$\,K.
	 \textit{Bottom}: Same histogram as in the center panel, for cores only.
  }
  \label{fig:Tb_info}
\end{figure}

\begin{figure}[t!]
  \centering
  \includegraphics[width=70mm]{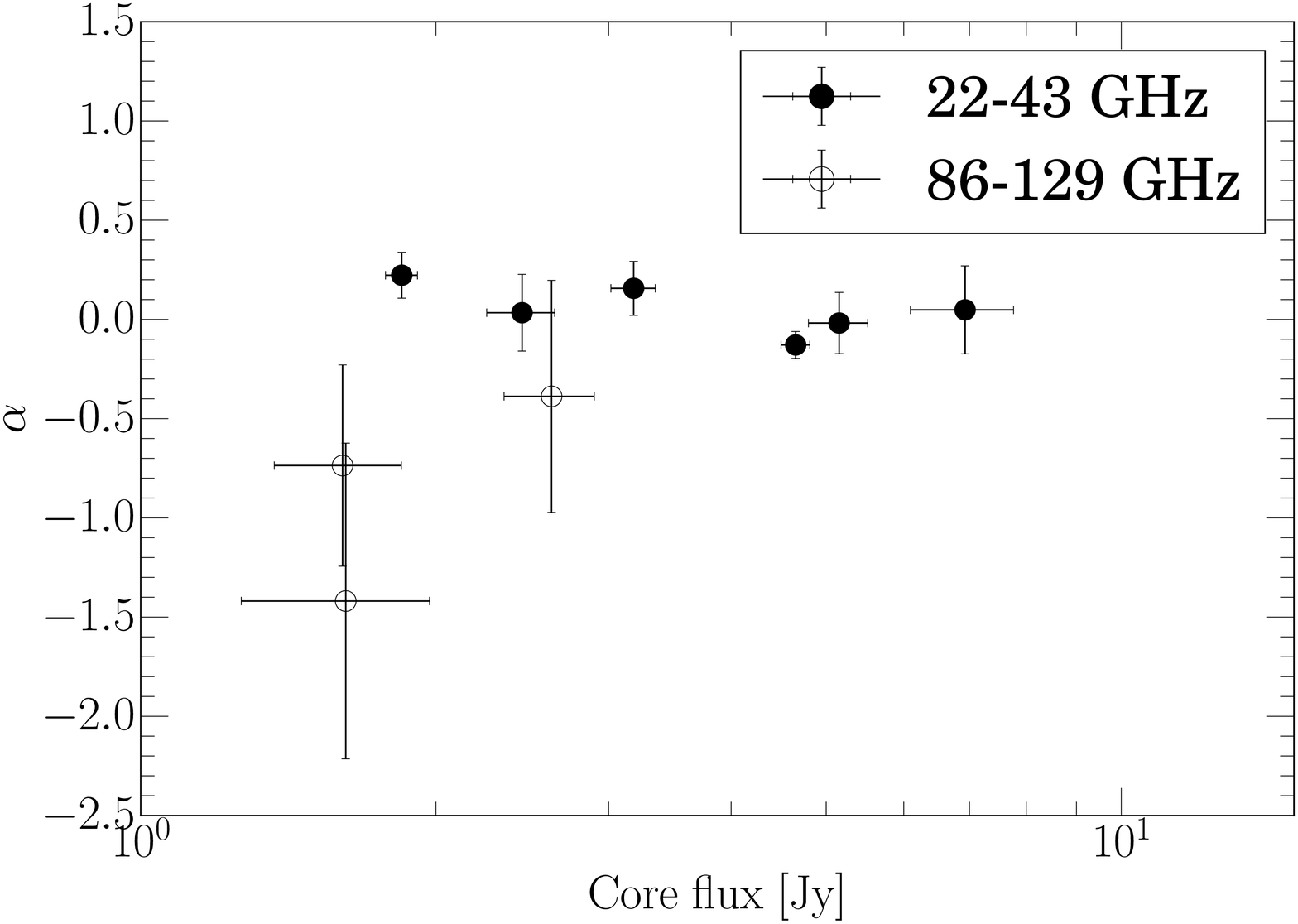}
  \includegraphics[width=73mm]{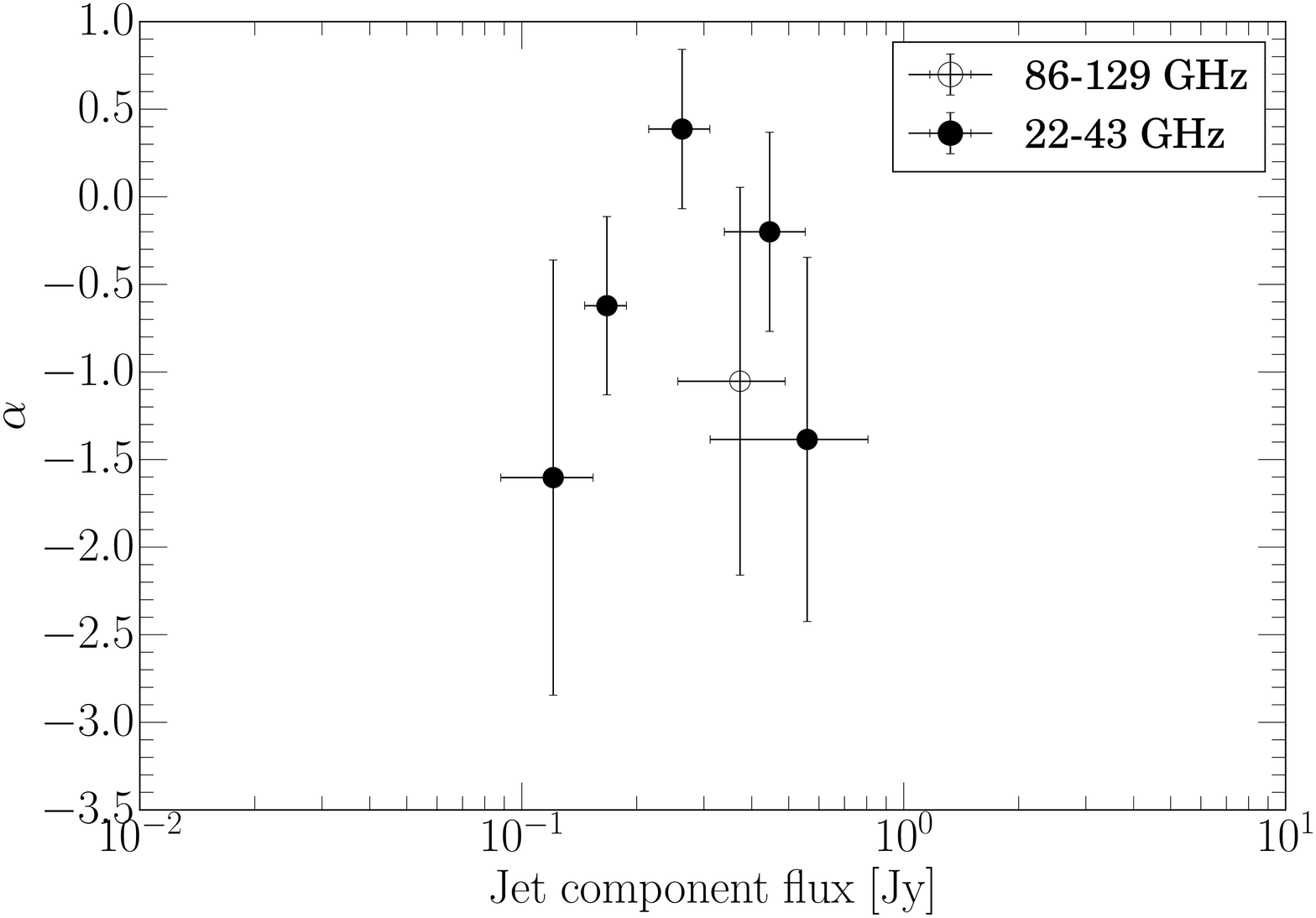}
  \includegraphics[width=70mm]{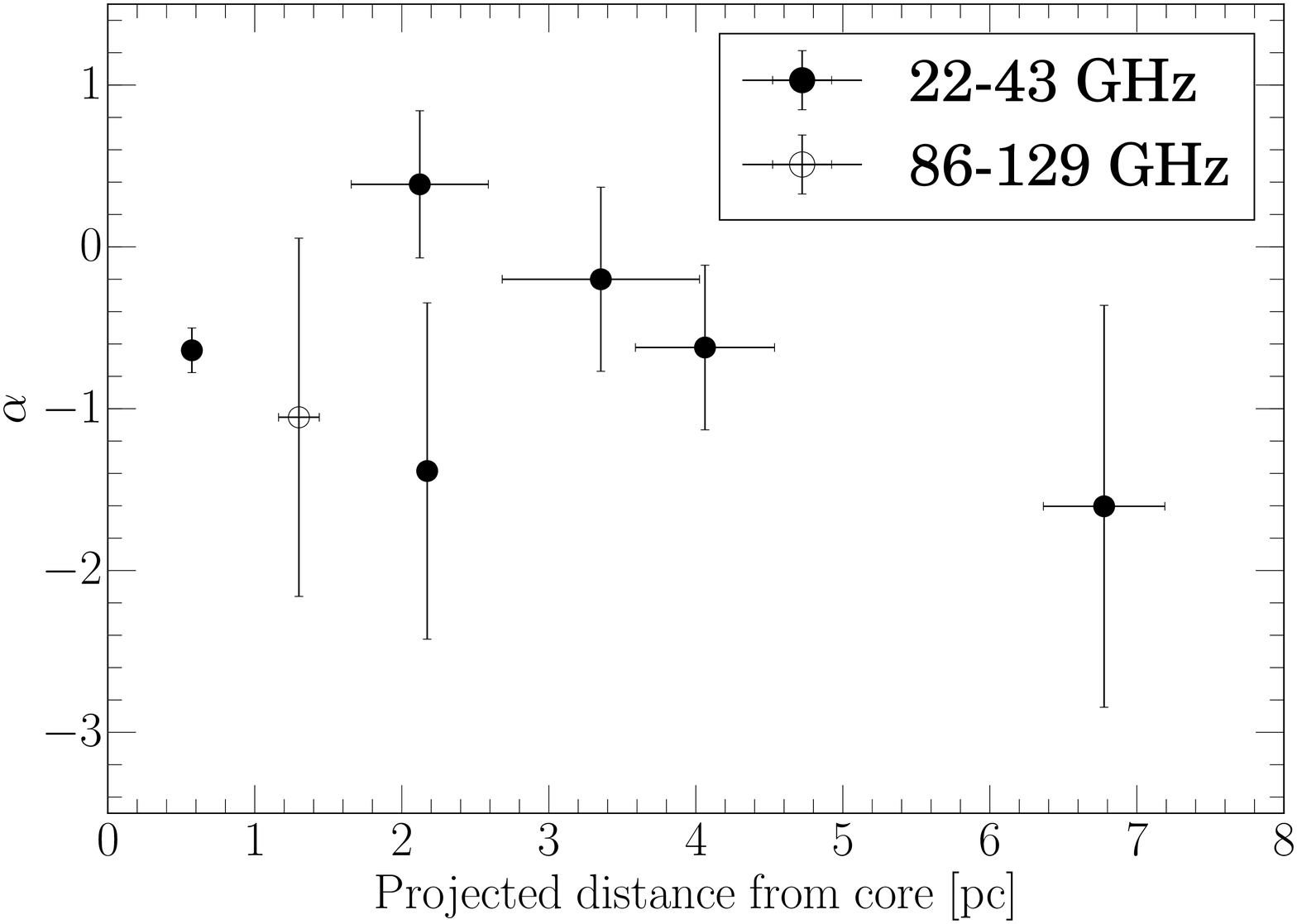}
  \caption{Spectral index distributions for Gaussian model components.
	 \textit{Top}: Spectral index of the cores for the frequency pairs 22--43 and 86--129 GHz as function of 22 and 86 GHz fluxes.
	 \textit{Center}: Same as top panel but for jet components.
	 \textit{Bottom}: Spectral index of jet components for the frequency pairs 22--43 and 86--129 GHz as function of projected distance from the core.
  }
  \label{fig:spix_info}
\end{figure}

\begin{figure}[t!]
  \centering
  \includegraphics[width=67mm]{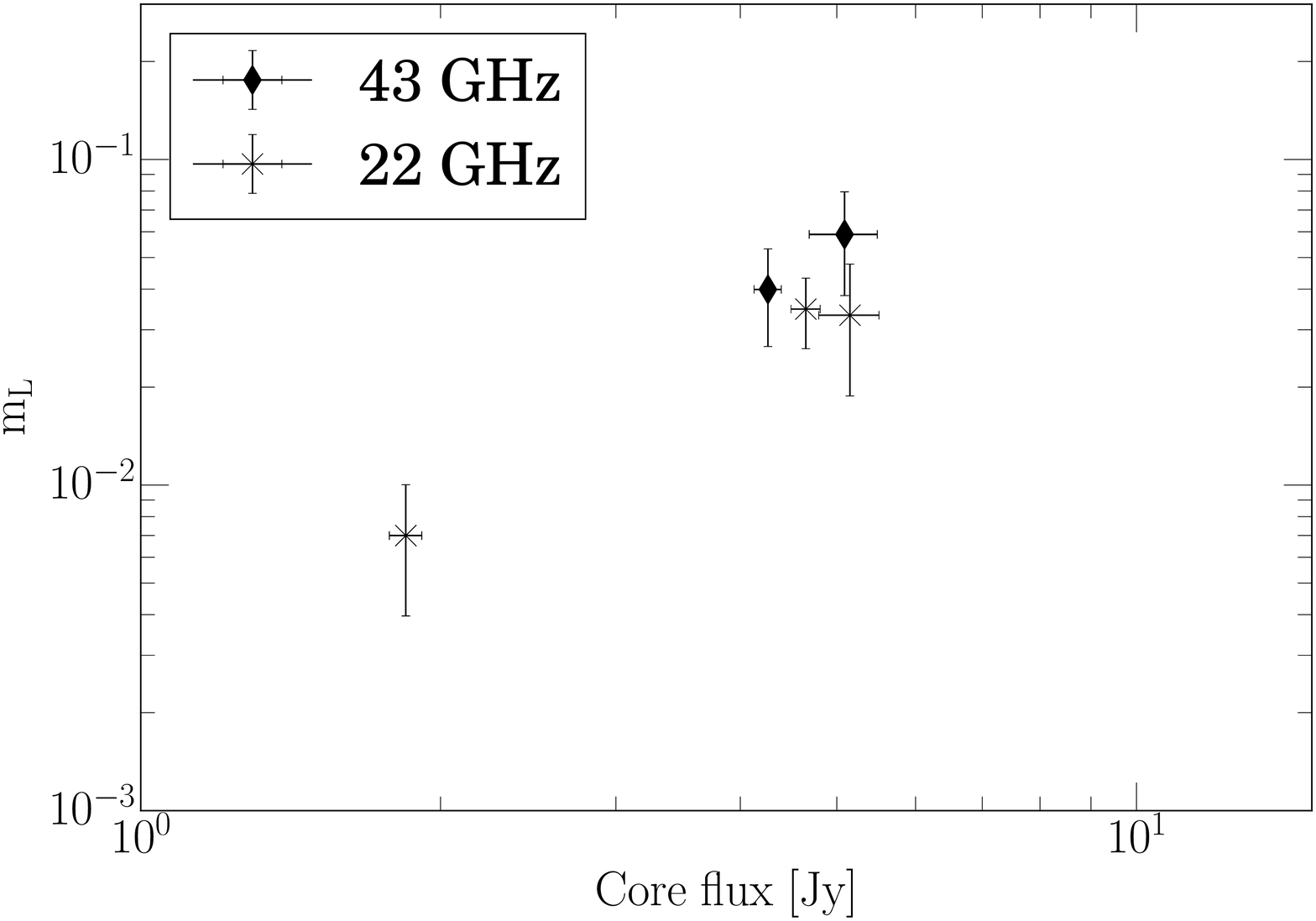}
  \includegraphics[width=70mm]{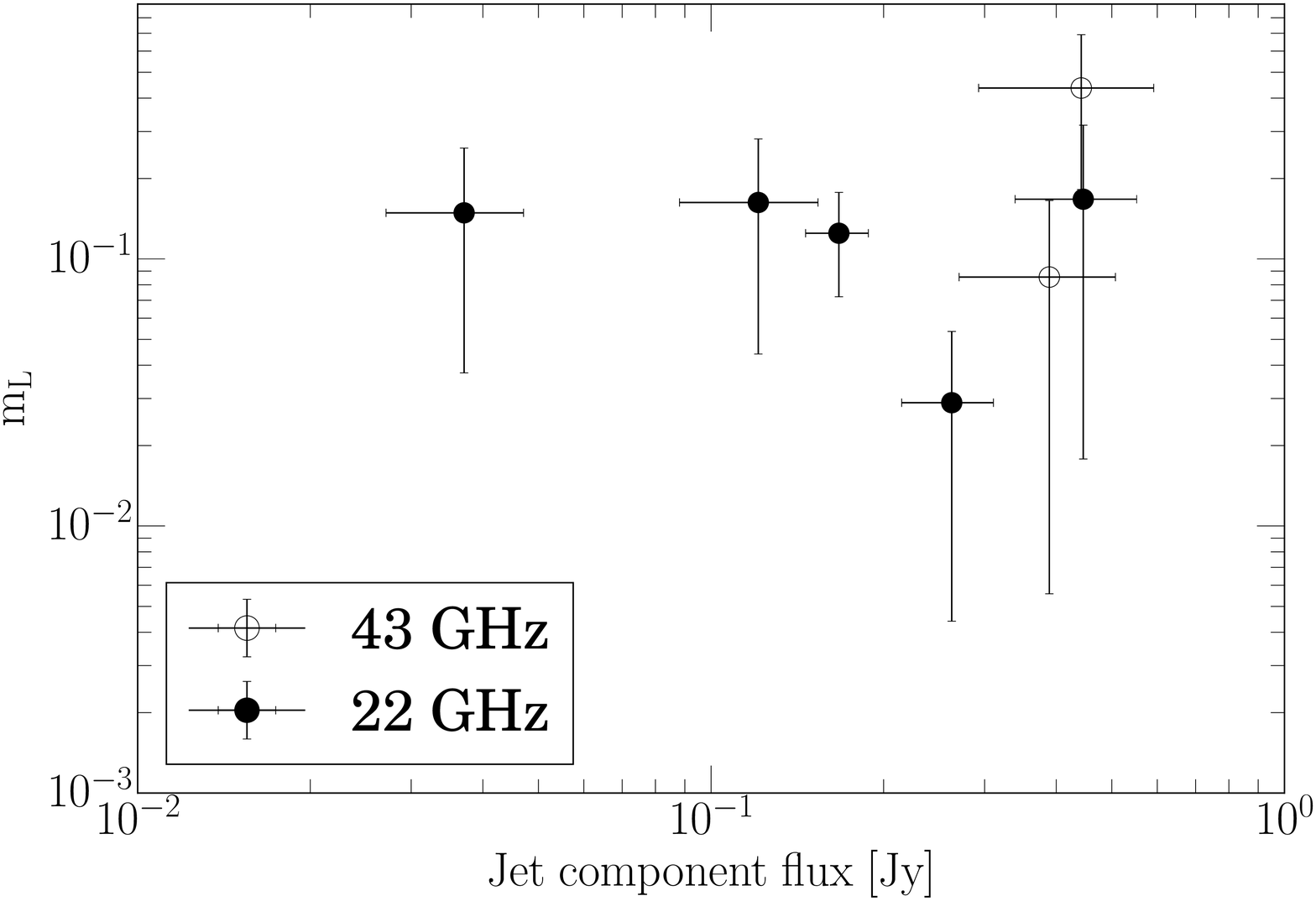}
  \includegraphics[width=70mm]{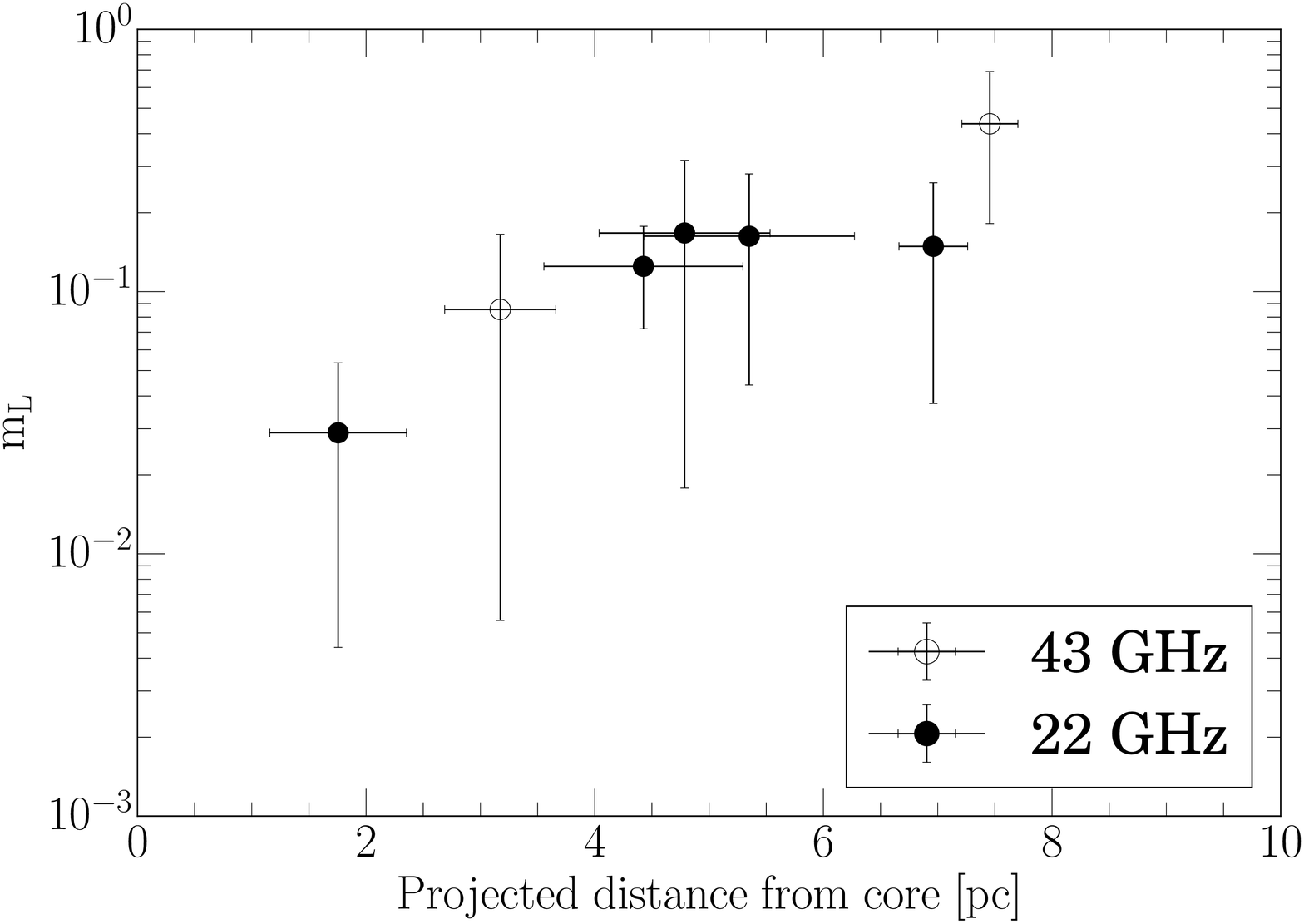}
  \caption{Polarization fractions of Gaussian model components, at 22 and 43 GHz.
  \textit{Top}: $m_{L}$ of cores versus flux.
  \textit{Center}: $m_{L}$ of jet components versus flux.
  \textit{Bottom}: $m_{L}$ of jet components versus projected distance from core.
  }
  \label{fig:pol_info}
\end{figure}

\subsection{Plasma Opacity and Degree of Polarization}

In order to understand the evolution of linear polarization as function of distance from the core (bottom panel of Figure~\ref{fig:pol_info}),
we investigate possible relations between the opacity of the relativistic plasma and the observed degree of linear polarization.
Theoretically, without Faraday rotation, the maximum degree of linear polarization $m_{\rm max}$ of the radiations from an ensemble of relativistic electrons in a uniform magnetic field 
can be related to the power law index $\gamma$ of the energy distribution of the relativistic electrons like 
\begin{equation}
  m_{\rm max, thin}= \frac{\gamma+1}{\gamma+7/3}= \frac{-\alpha +1}{-\alpha +5/3}   \label{eq:thinpol}
\end{equation}
\citep{pacholczyk70} where $m_{\rm max,thin}$ is the maximum degrees of linear polarization for optically thin plasmas,
$\gamma$ is the index for a power-law energy distribution of the electrons (i.e. $\gamma>0$ for steep energy spectrum), and
$\alpha$ is the spectral index.
We note that the second equality of Equation (\ref{eq:thinpol}) assumes $\alpha=-(\gamma-1)/2$, which is only valid for an optically thin synchrotron emission region (i.e., $\alpha<0$).

\begin{figure}[t!]
  \centering
  \includegraphics[trim=0mm 3mm 0mm 2mm, clip, width=70mm]{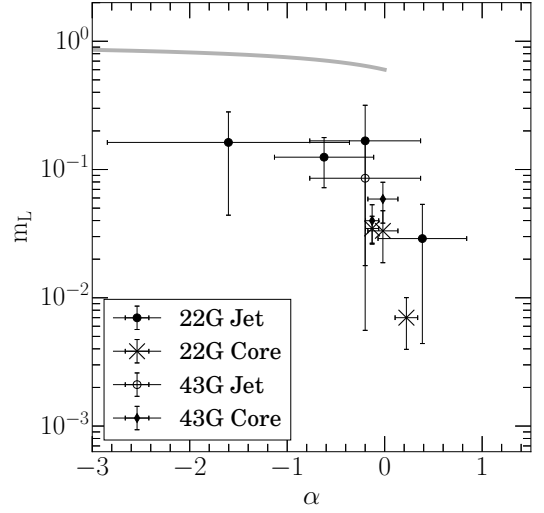}
  \caption{Degree of linear polarization $m_{L}$ as function of spectral index $\alpha$ for cores and jet components for which polarization is detected. 
  Gray curve shows the maximum degrees of polarization expected for optically thin and thick plasma
  (Equation~(\ref{eq:thinpol})).
  }
  \label{fig:pol_info2}
\end{figure}

Figure \ref{fig:pol_info2} shows the degree of observed linear polarization $m_L$ as function of spectral index $\alpha$.
The general trend found in our data is consistent with the theoretical expectation;
the observed degree of polarization is larger for components with steeper spectra, although we note that the basic assumption of optically thin synchrotron emitters breaks down for the data with $\alpha \gtrsim 0$. Even though, the overall trend in the observed data implies that the observed variations of linear polarization are related to the internal evolution of the pc-scale jets, 
notably decreasing particle densities associated with the expansion of jet components after ejection from the core. 
We note that our observed values of $m_{L}$ as function of $\alpha$ are much smaller than the theoretical maxima (e.g. $\sim$70\% for $\alpha=-0.5$ according to Equation~(\ref{eq:thinpol})), 
in agreement with previous mm-wavelength polarimetric AGN surveys (e.g., \citealt{jorstad07,trippe10,trippe12b} and references therein). 
This indicates that the underlying assumption of an idealized homogeneous, isotropic synchrotron plasma is oversimplified.

A plausible explanation for the low intrinsic degrees of polarization is provided by randomized magnetic field configurations inside the outflows.
Assuming that a radio jet consists of multiple magnetized turbulent plasma cells with uniform magnetic field strength but with randomly varying orientation, the intrinsic linear polarization will cancel out partially due to its vectorial nature when observed with finite angular resolution.
Following \cite{hughes91}, the relation between intrinsic (e.g., maximum $m_{L}$ given by Equation~(\ref{eq:thinpol})) and observed linear polarization is
\begin{equation}
  m_{L} \approx m_{\rm max}/\sqrt{N} \pm m_{\rm max}/\sqrt{2N}
  \label{eq:pol_depol}
\end{equation}
where $N$ is the number of turbulence cells within the telescope beam; $N$ provides a measure of the degree of turbulence in a given AGN outflow.
We calculated $N \simeq m^{2}_{\rm max}/m^{2}_{L}$ considering only components with $\alpha<-0.1$ because Equation (\ref{eq:thinpol}) is based on the assumption of optically thin synchrotron emitters.
From this, we found a minimum value of $N_{\rm min}=14$, a maximum value of $N_{\rm max}=327$, and an average value of $N_{\rm mean}=117$ (with the standard deviation being 123).
Accordingly, the radio jets in our sample seem to be highly turbulent.

Although the turbulence model provides a satisfactory explanation for observed weak polarizations ($\lesssim$10\%), 
it does not explain how some of the jet components maintain strong polarization; the most extended jet component of BL Lac at 43 GHz is polarized up to $\sim$ 40\% with high signal to noise (see the $m_{L}$ map in Figure~\ref{fig:bllac_polmaps}).
Such strong polarization requires small values of $N$, meaning a rather uniform magnetic field structure over spatial scales on the order of, or larger than, the beam size.

\begin{figure}[t!]
  \centering
  \includegraphics[width=55mm]{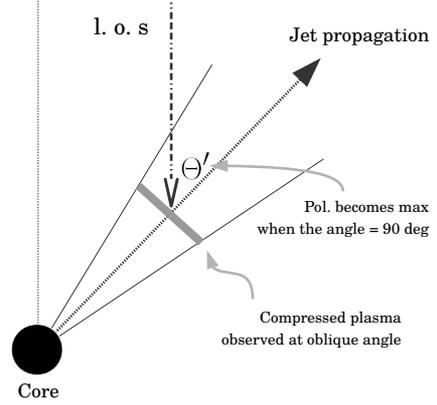}
  \caption{Geometry of a transverse shock traveling downstream along a radio jet seen from the shock frame, as assumed for Equation~(\ref{eq:mpol_shock}). The degree of `repolarization' depends on shock strength and viewing angle.}
  \label{fig:ShockGeometry}
\end{figure}

It is known that shearing or compression of the plasma and the magnetic field lines frozen therein can change levels of polarization and polarization position angles (e.g., \citealt{laing80,laing08}).
As for compressing the plasma, shocks in the parsec-scale jets may be especially important since they (i) provide higher degree of compression, (ii) work on shorter timescales due to their speed faster than the sound speed of the gas, and (iii) lead to the release of large amounts of energy - usually in the form of flares.
In the light of this, we investigate in the following whether an internal shock traveling downstream of the jet of BL Lac could reproduce such high $m_{L}$ by compressing turbulent plasma.
After compression by the shock, initially random magnetic field lines still appear randomly oriented when the shock front is observed face-on but will assume an ordered configuration in projection when observed edge-on.
The `repolarization' depends on the characteristics of the shock front such as shock strength and angle with respect to the line of sight.
Detailed calculations for a shock front oriented transverse to the jet direction show that the level of polarization for an optically-thin relativistic jet component is
\begin{eqnarray}
  m_{L} & \approx & \frac{-\alpha+1}{-\alpha+5/3} \times  \frac{(1-\eta^{-2}\sin^{2}\Theta')}{2-(1-\eta^{-2})\sin^{2}\Theta'} \label{eq:mpol_shock} \\
  \eta & = & n_{\rm shocked}/n_{\rm unshocked} \label{eq:eta_shock} \\
  \Theta' & = & \tan^{-1}\frac{\sin \Theta}{\Gamma\left(\cos \Theta - \sqrt{1-\Gamma^{-2}}\right)} \label{eq:angle_shock}
\end{eqnarray}
\citep{hughes85,hughes91} where the first term of Equation~(\ref{eq:mpol_shock}) is simply Equation~(\ref{eq:thinpol}),
$\eta$ is the ratio of the particle densities $n$ in shocked and unshocked plasma, respectively,
$\Theta'$ is the jet viewing angle corrected for relativistic aberration (i.e. the viewing angle in the shock frame),
$\Gamma$ is the bulk Lorentz factor of a traveling jet component, and
$\Theta$ is the viewing angle of the jet component as seen in the reference frame of observer.
Figure \ref{fig:ShockGeometry} illustrates the geometry considered here.

\begin{figure}[t!]
  \centering
  \includegraphics[trim=3mm 3mm 3mm 1mm, clip, width=80mm]{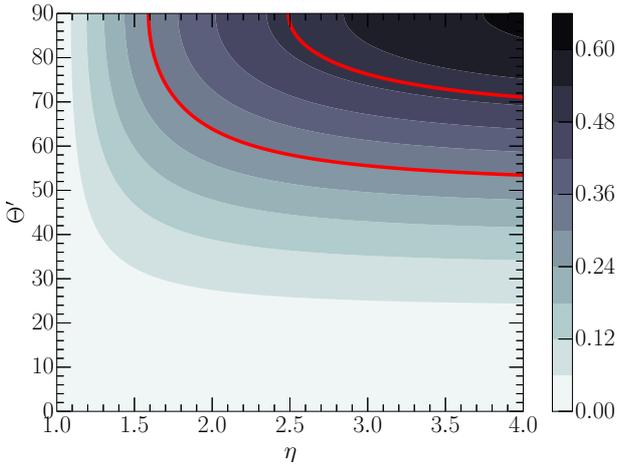}
  \caption{Map of polarization fraction $m_{L}$ (color bar) as function of $\eta$ and $\Theta'$ according to Equation~(\ref{eq:mpol_shock}). The red contour lines correspond to $m_{L}=0.3$ and 0.5, respectively.}
  \label{fig:HughesShockModel}
\end{figure}

In order to determine the shock strength, we need to know $\Theta$ and $\Gamma$ which can be derived from time-resolved monitoring of individual jets. 
As our current analysis is limited to single-epoch observations, we rather probe which combinations of the parameters $\eta$ and $\Theta'$ lead to polarization levels in agreement with observation..
Assuming an optically thin jet component with $\alpha=-0.5$, we calculate the shock-modified $m_{L}$ by following Equation~(\ref{eq:mpol_shock}).
The result is shown in Figure~\ref{fig:HughesShockModel}.
In order to explain the observed $m_{L}$, the shock strength $\eta$ needs to be at least $\gtrsim1.5$)
and the shock plane should be close to parallel to the line of sight in the shock frame ($\Theta'\gtrsim60^{\circ}$) (albeit absolute EVPA information
is required to examine whether the shock and the orientation of the compressed magnetic field are indeed perpendicular to the jets as the above discussion assumes; see Figure~13 of \citealt{jorstad07} for an example). 
Accordingly, regular monitoring observations of $m_{L}$ and absolute EVPA, at multiple frequencies, should be performed in forthcoming observations to fully understand the evolution of AGN radio jets coupled to large-scale magnetic fields.

\section{Summary\label{sec:summary}}

In this paper we presented results of first-epoch KVN observations conducted in the frame of the \emph{P}lasma Physics of \emph{A}ctive \emph{Ga}lactic \emph{N}uclei (PAGaN) project, aimed at exploring spectral and linear polarization properties of parsec-scale radio jets of AGN.
We observed seven sources at 22, 43, 86, and 129 GHz in dual polarization.
The main conclusions of our study are as follows.

\begin{enumerate}
  
  \item We found brightness temperatures $T_{\rm b,app}$ of $10^{8.76 \pm 0.98}$\,K for resolved jet components and obtained the lower limits for $T_{\rm b,app}$ of the other core and jet components as $10^{9.86}$\,K and $10^{8.01}$\,K, respectively. 
        Although the measurements are limited by the instrumental angular resolution, 
        these values imply relativistic kinematics of the observed radio jets. 
        
  \item We found the fractional linear polarization of jet components to increase with larger projected distance from the core. 
	 This indicates either that an external depolarizing medium surrounds the jet or that the fractional polarization $m_{L}$ is directly related to the internal evolution of jet components while they are propagating outward.
	 When analyzing $m_L$ as function of spectral index $\alpha$, we found that $m_L$ is systematically higher for emitters with smaller $\alpha$, as expected for optically thin synchrotron plasmas. 
	 This implies that the evolution of $m_L$ is driven by the evolution of the plasma opacity: the further outward the component, the more it expands and the more transparent it becomes, leading to increasing polarization.
	 
  \item We analyzed the polarization levels, $m_L\lesssim10$\%, found in the observed jets by assuming 
         (a) turbulence in the jets with $N$ randomly oriented polarization cells inside the observing beam size and (b) transverse shocks propagating downstream along the jets and compressing the magnetized plasma. 
	 Our calculations suggest that the jets are indeed turbulent, with 
	 $N_{\rm min}=14$, $N_{\rm max}=327$, and $N_{\rm mean}=117$.
	 We specifically applied the shock model to a strongly polarized jet component of BL Lac ($m_{L} \sim 40\%$ at 43 GHz).
	 We were able to constrain the range of values permitted for compression factor and viewing angle to $\eta\gtrsim1.5$ and $\Theta' \gtrsim 60^{\circ}$, respectively. 
	 This indicates that shocks reducing the polarization to the observed values can be of various strengths, with the shock front being close to parallel to the line of sight.

\end{enumerate}

The PAGaN project is ongoing, with further KVN polarimetric and KaVA total intensity observations being conducted.
In the near future, KaVA polarimetry is expected to become available at 22~GHz, and KVN polarimetry at 86 and 129 GHz to become more stable. With the advance of VLBI networks in East Asia and with forthcoming polarimetric time-series monitoring observations being prepared, it will be possible to probe the nature of radio jets in active galaxies in more detail.


\acknowledgments

This project has been supported by the Korean National Research Foundation (NRF) via Basic Research Grant 2012-R1A1A-2041387.
We are grateful to all staff members of KVN who helped to operate the array and to correlate the data. The KVN is a facility operated by the Korea Astronomy and Space Science Institute. The KVN operations are supported by KREONET (Korea Research Environment Open NETwork) which is managed and operated by KISTI (Korea Institute of Science and Technology Information).


\appendix

\end{document}